\documentclass[structabstract]{aa}  
\usepackage{graphicx}
\usepackage{txfonts}
\usepackage{natbib}
\newcommand{\teff}  {T$_\mathrm{eff}$}
\newcommand{\logg}  {$\log g$}

\begin{document}
   \title{Li depletion in solar analogues with exoplanets}

   \subtitle{Extending the sample\thanks{
Based on observations collected at the La Silla Observatory, ESO
(Chile), with the HARPS spectrograph at the 3.6 m ESO telescope, 
with CORALIE spectrograph at the 1.2 m Euler Swiss telescope and 
with the FEROS spectrograph at the 1.52 m ESO telescope;
at the Paranal Observatory, ESO (Chile), using the UVES spectrograph
at the VLT/UT2 Kueyen telescope, and with the FIES, SARG and UES
spectrographs at the 2.5 m NOT, the 3.6 m TNG and the 4.2 WHT, respectively, 
both at La Palma (Canary Islands, Spain).
}}

   \author{E. Delgado Mena\inst{1}
      \and G. Israelian\inst{2,3} 
      \and J.~I.~Gonz\'alez Hern\'andez\inst{2,3}
      \and S. G. Sousa\inst{1,2,4}
      \and A. Mortier\inst{1,4}
      \and N.~C.~Santos\inst{1,4}
      \and V. Zh. Adibekyan\inst{1}
      \and J. Fernandes\inst{5}
      \and R. Rebolo\inst{2,3,6}
      \and S. Udry\inst{7}
      \and M. Mayor\inst{7}
      }
\institute{
Centro de Astrof\'isica, Universidade do Porto, Rua das Estrelas,
4150-762, Porto, Portugal 
              \email{Elisa.Delgado@astro.up.pt}
\and 
Instituto de Astrof\'{\i}sica de Canarias,
C/ Via Lactea, s/n, 38200, La Laguna, 
Tenerife, Spain 
\and 
Departamento de Astrof\'isica, Universidad de La Laguna, 38205 La Laguna, Tenerife, Spain
\and
Departamento de F\'isica e Astronomia, Faculdade de Ci\^encias, Universidade do Porto, Portugal
\and
CGUC, Department of Mathematics and Astronomical Observatory, University of
Coimbra, Portugal 
\and  
Consejo Superior de Investigaciones Cient\'{\i}ficas, Spain
\and  
Observatoire de Gen\`eve, Universit\'e de Gen\`eve, 51 ch. des Maillettes, 1290 Sauverny,
Switzerland 
}     


   \date{Received ...; accepted ...}

 
  \abstract
{}
{We want to study the effects of the formation of planets and planetary systems on the atmospheric Li abundance of planet host stars.}
{In this work we present new determinations of lithium abundances for 326 Main Sequence stars with and without planets in the \teff\ range 5600-5900 K. 277 stars come from the HARPS sample, the remaining targets have been observed with a variety of high resolution spectrographs.}
{We confirm significant differences in the Li distribution of solar twins (\teff\ = T$_{\odot} \pm$ 80 K, \logg\ = \logg$_{\odot}$ $\pm$ 0.2 and [Fe/H] = [Fe/H]$_{\odot} \pm$ 0.2):
the full sample of planet host stars (22) shows Li average values lower than "single" stars with no detected planets (60). 
If we focus in subsamples with narrower ranges in metallicity and age, we observe indications of a similar result though it is not so clear for some of the studied subsamples. Furthermore, we compare the observed spectra of several couples of 
stars with very similar parameters which show different Li abundances up to 1.6 dex. Therefore we show that neither age, nor mass nor metallicity of a parent star is the only responsible for enhanced Li depletion in solar analogues.} 
{We conclude that another variable must account for that difference and suggest that this could be the presence of planets which causes additional rotationally induced mixing in the external layers of planet host stars. Moreover, we find indications that the amount of depletion of Li in planet host solar-type stars is higher when the planets are more massive than Jupiter}


   \maketitle
%

\section{Introduction}

The study of extrasolar planets has been an exciting field of astrophysics for more than 15 years. More than 1000 planets are known in almost 800 planetary systems (Encyclopaedia of Extrasolar Planets, \citet{schneider}). Photospheric abundances of planet host stars are key to understand the role of the metallicity of protoplanetary clouds in the formation of planets and to determine how the formation of planets may affect the structure and evolution of the parent stars.\\

Many studies \citep[e.g.][]{gonzalez97,santos04,santos05,fischer,sousa08} showed the significant metallicity excess of the planet-host sample compared with the sample of stars without known giant planets, suggesting  a possible clue for a formation scenario. However, this metallicity excess does not seem to appear in stars which only host Neptunians and Earth-like planets \citep{udry07,sousa08,sousa_harps2}. These stars have metallicities lower on average than stars with Jupiters, and more similar to the [Fe/H] distribution of stars without detected planets. Interestingly, \citet{adibekyan12a,adibekyan12b} recently showed that even low-mass planet hosts with [Fe/H] $<$ -0.2 present enhanced abundances of $\alpha$ elements and suggested that a minimum quantity of refractory material is required to form planets when the amount of Fe is low. \\ 

\begin{figure}[h!]
\centering
\includegraphics[width=6.7cm,angle=270]{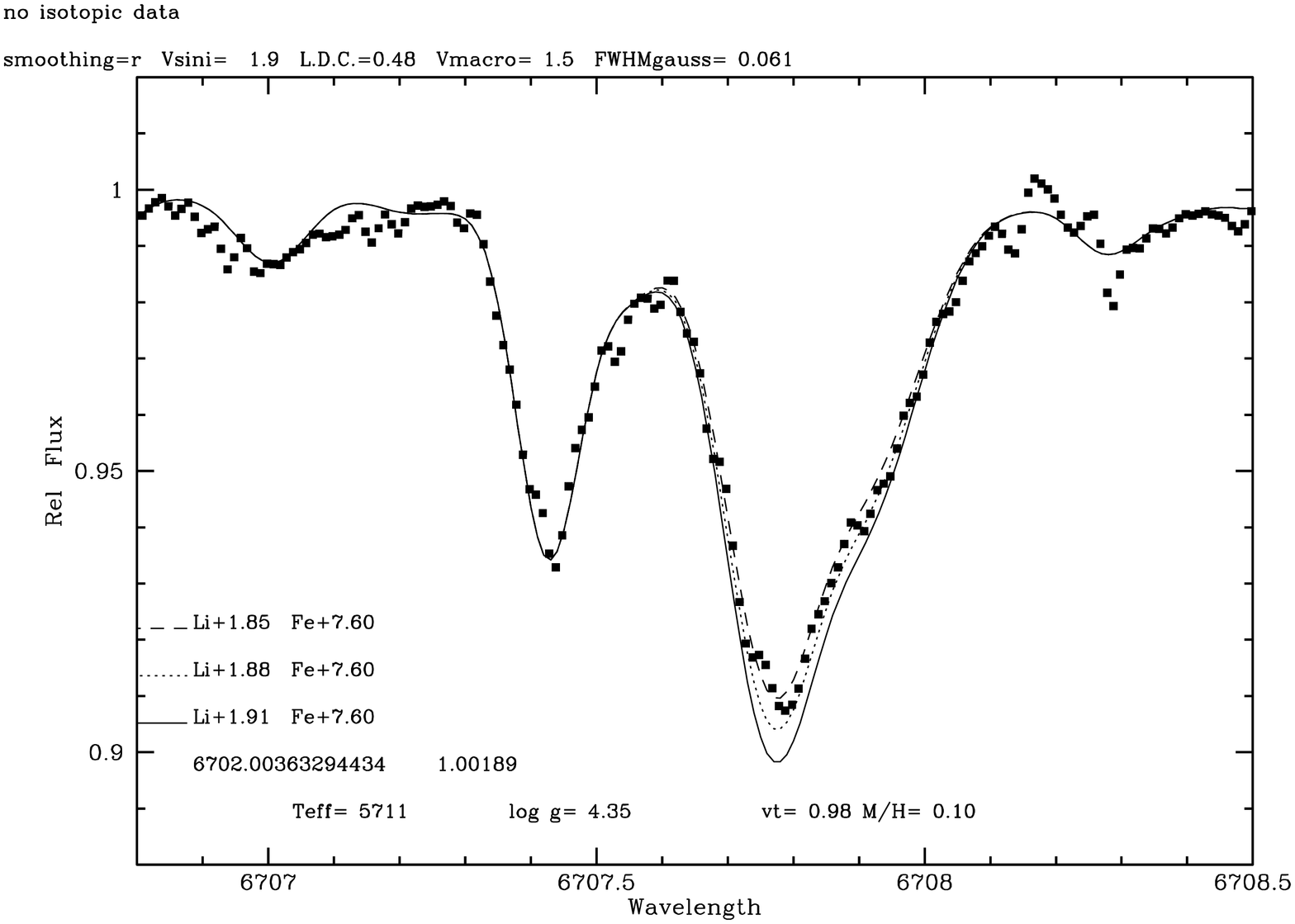}
\includegraphics[width=6.7cm,angle=270]{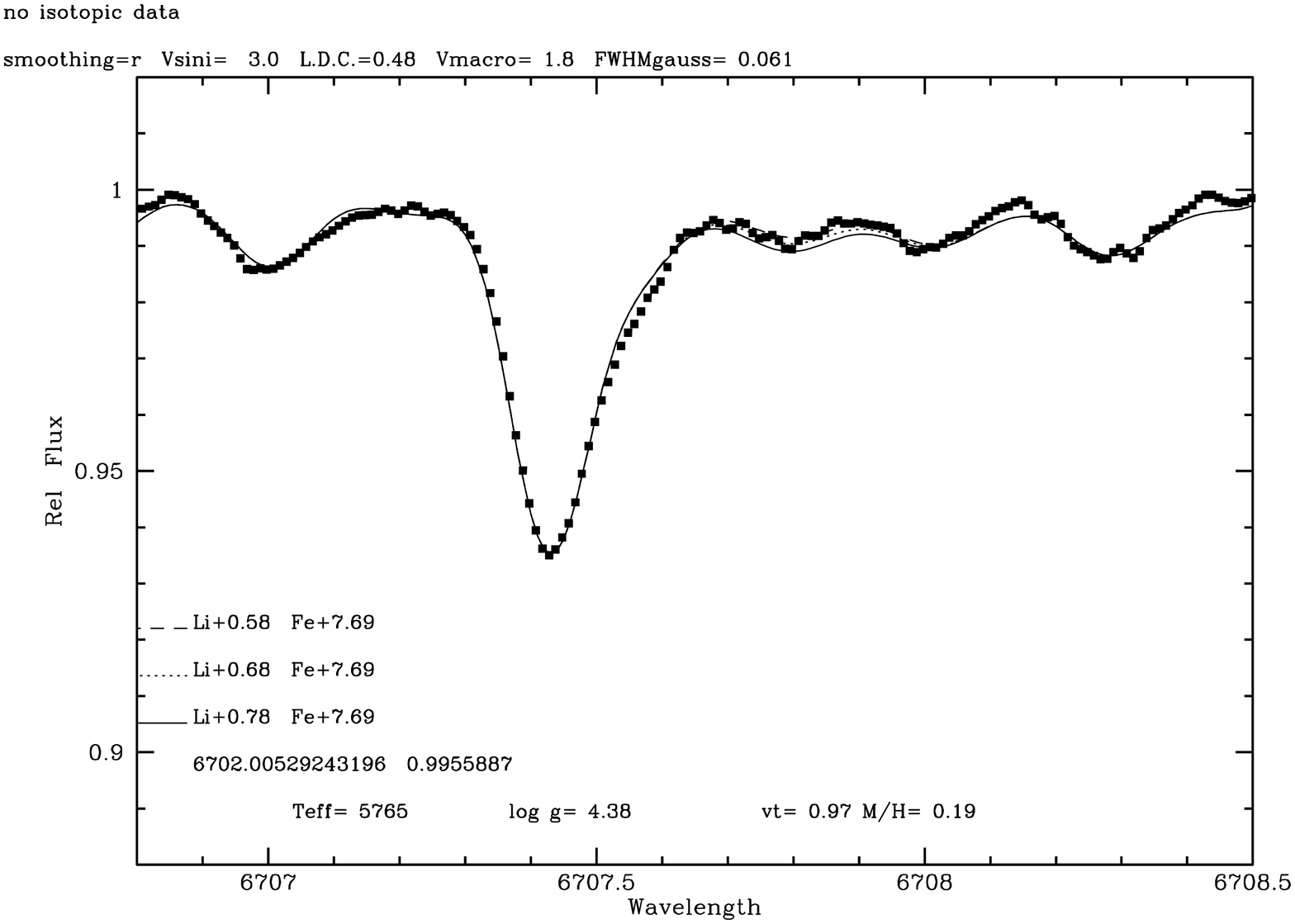}
\includegraphics[width=6.7cm,angle=270]{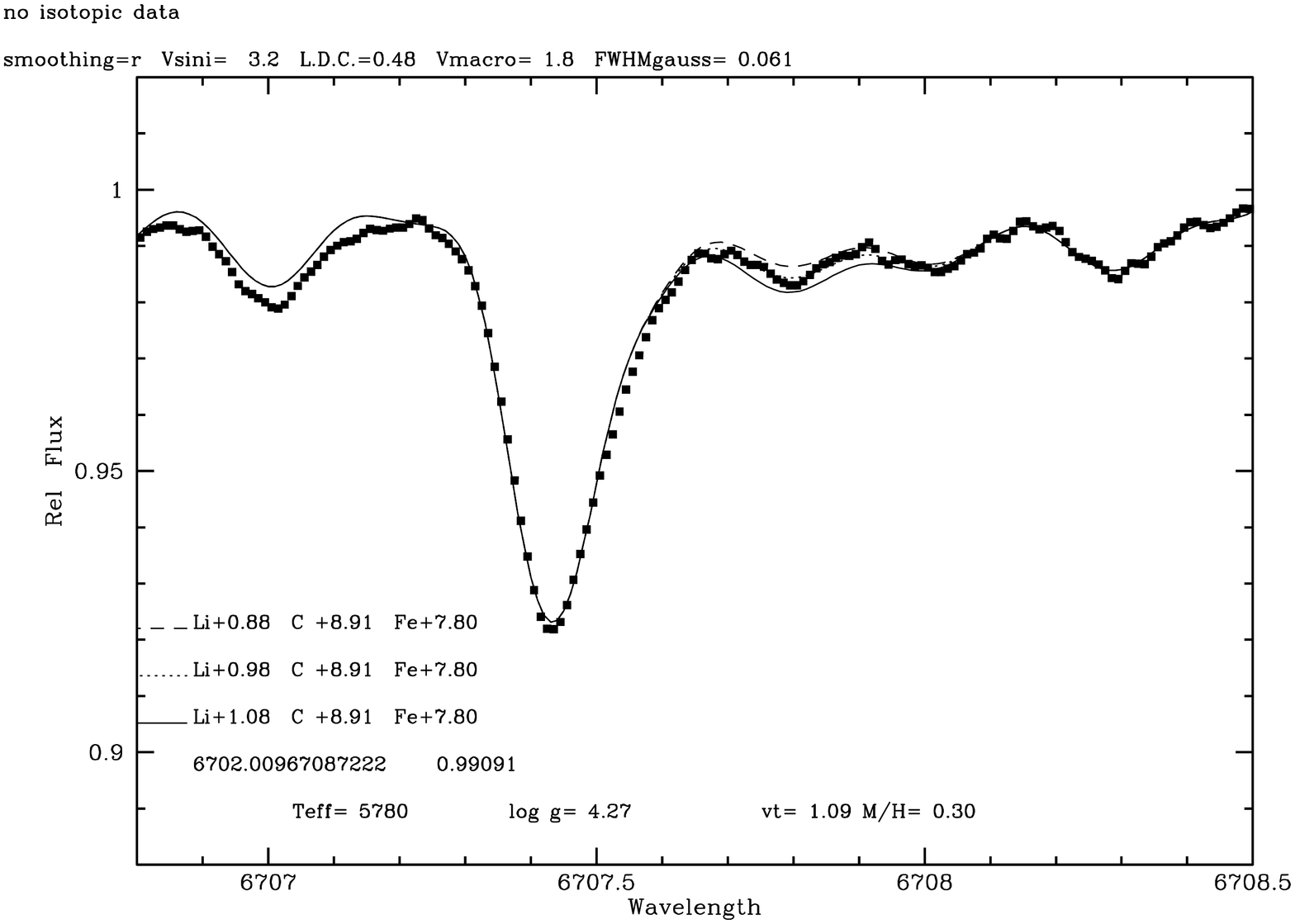}

\caption{Spectral synthesis around Li region for the stars HD96423, HD1461 and HD160691.} 
\label{Li_synthesis}
\end{figure}

\begin{center}
\begin{table*}
\caption{Observing details: telescopes, spectrographs, resolving
power and spectral ranges used in this work. }
\centering
\begin{tabular}{lccc}
\hline
\hline
\noalign{\smallskip}
telescope & instrument & resolution & spectral range\\
 & & $\lambda/\delta\lambda $ & \AA \\
\noalign{\smallskip}
\hline    
\noalign{\smallskip}
3.6-m ESO La Silla Observatory (Chile) & HARPS & 100,000 & 3.800--7.000\\
8.2-m Kueyen UT2 (VLT) & UVES & 115,000 & 3.000--4.800,4.800--6.800 \\
2.2-m ESO/MPI telescope & FEROS & 48,000 & 3.600--9.200 \\
3.5-m TNG & SARG & 57,000/86,000 & 5.100--10.100\\
2.6-m Nordic Optical Telescope & FIES & 67,000 & 3.700--7.300\\
1.93-m OHP & SOPHIE& 75,000 & 3.820--6.930 \\       
1.2-m Euler Swiss telescope & CORALIE & 50,000 & 3.800--6.800\\
4.2-m William Herschel Telescope & UES & 55,000 & 4.600--7.800 \\
\noalign{\smallskip}
\hline 
\end{tabular}
\label{tabinstr}
\end{table*}      
\end{center}

\citet{king97} reported a difference in the Li abundances between the components of the binary 16 Cyg suggesting that the presence of a planet might be responsible for the lower Li found in its host star. Using a larger sample of stars in the \teff\ range 5600-5850K with detected planets, \citet{gonzalez00} and \citet{israelian04} showed that exoplanets hosts are significantly more Li-depleted than stars without detected planets (hereafter called "single" stars\footnote{We adopt this term for the sake of simplicity along the paper. We warn the reader not to mix up with the term tipically given to non-binary stars.}). This result was also supported by \citet{takeda05}, \citet{chen06} and \citet{gonzalez_li08}. 
\citet{israelian09} confirmed former results using the homogeneous and high quality HARPS GTO sample. They found that about 50\% of 60 solar analogues (\teff\ = T$_{\odot}$\footnote{T$_{\odot}$=5777 K} $\pm$ 80 K) without detected planets had A(Li) $\ge$ 1.5 while only 2 out of 24 planet hosts had high Li abundances. Other recent works by \citet{takeda10,gonzalez_li10} have also reported similar results. Furthermore, \citet{sousa_li} showed that mass and age are not responsible for the observed correlation using the same sample as in \citet{israelian09}. 
Doubts about the proposed Li-planet connection have been raised in several works \citep{ryan00,luck06,baumann,ghezzi_li,ramirez_li12} arguing that the observed behaviour of Li abundances could be associated to age, mass or metallicity differences between stars with and without planets. However, this cannot be the ultimate explanation of the observed behaviour of Li in planet host stars, as several old stellar clusters like M67 have shown that solar-type stars of very similar age and metallicity present a wide dispersion of Li abundances \citep{randich07,pasquini08,pace12}.
So, what is the principal parameter that accounts for this scatter in Li abundances?\\

Lithium is formed in a significant quantity at the primordial
nucleosynthesis and is easily destroyed by (p, $\alpha$)-reactions at 2.5 million K in the inner layers of solar-type stars. Although Li depletion occurs primarily in the pre-Main Sequence (PMS), it can also take place in stellar envelopes if any extra mixing process exists. In fact, there is an obvious relation between \teff~and Li-depletion: cooler stars are
more Li-depleted because of their thicker convective envelopes (where
Li is brought to hotter layers to be depleted).\\  

It seems that rotation-induced mixing and angular momentum loss are
the most efficient processes destroying Li in solar-type stars on the Main Sequence (MS)
\citep[e.g.][]{zahn92,pinsonneault92,deliyannis97}.
Observations indicate that rapidly rotating stars preserve more Li than slow rotators of the same
mass as observed in Pleiades \citep{soderblom93,garcialopez} or IC 2602 
\citep{randich97}. This is also found in solar-type stars by \citet{takeda10} 
and \citet{gonzalez_li10}. Other invoked mechanisms are internal waves \citep[e.g.][]{montalban96} or overshooting mixing, recently claimed as being a primary depletion mechanism \citep[e.g.][]{xiong09,zhang12}. Some works have also combined several mechanisms showing that the efficiency of rotational mixing is decreased when magnetic fields are taken into account \citep{eggenberger10} as well as when there are internal gravity waves \citep{charbonnel05} since they lead to efficient angular momentum redistribution from the core to the envelope.\\

Other mechanisms directly related to the presence of planets have been proposed to cause additional Li depletion. 
For instance, \citet{israelian04}, \citet{chen06} and \citet{castro09} 
suggested that Li depletion could be related to planet migration since it
can create a shear instability that produces effective mixing. It is
also possible that proto-planetary discs lock a large amount of
angular momentum and therefore create some rotational breaking in the
host stars during the PMS inducing an increased mixing
\citep{israelian04}. This possibility has also been
proposed by \citet{bouvier08} using simple rotational models:
long-lived accretion discs of stars with small initial velocities
lead to the formation of a large velocity shear in the base of the
convective zone. The velocity gradient in turn triggers
hydrodynamical instabilities responsible for enhanced lithium burning
on PMS and MS evolution scales. This explanation matches well the Li 
depletion in planet hosts since we expect long disc lifetimes in order 
to form planets. The relation of long-lived discs with slow rotation on
the ZAMS has also been probed by \citet{eggenberger12} as well as the 
increase of rotational mixing (and thus Li depletion) when differential 
rotation rises in the stellar interior during PMS.
Finally, the infall of planetary material might also affect the mixing processes 
of those stars by thermohaline convection \citep{garaud,theado12} as well as the episodic 
accretion of planetary material can increase the temperature in the bottom of the convective 
envelope and hence increase Li depletion \citep{baraffe10}.\\

The above mentioned theoretical studies and models propose a clear relationship between stellar rotation (pre MS or MS), formation and evolution of planets and surface Li abundances of solar type stars. To discover this relation observationally, homogeneous and precise studies of planet host stars have to be undertaken.In this paper we will extend our previous Li work \citep{israelian09} by including
new stars observed in HARPS surveys and at other telescopes.

\begin{center}
\begin{table*}
\caption{Definition of the subsamples in this work}
\centering
\begin{tabular}{lccccccc}
\hline
\hline
\noalign{\smallskip}
Name & \teff\ & [Fe/H] & \logg\ & planet hosts\tablefootmark{a} & single stars & planet hosts  & single stars \\
     & (K)  &   &     (cm\,s$^{-2}$) & & & A(Li)$>$1.4  & A(Li)$>$1.4 \\
\noalign{\smallskip}
\hline    
\noalign{\smallskip}
Solar type & 5600,5900 & -0.85,0.5 & 3.8,4.7 & 43+49 & 233\tablefootmark{b}& 28\% +5\% -4\% & 43\% $\pm$3\% \\
Solar analogues & 5697,5857 & -0.6,0.5 & 4.24,4.64 & 18+25 & 99& 19\%  +7\% -4\% & 47\% $\pm$ 5\% \\
Solar twins & 5697,5857 & -0.2,0.2 & 4.24,4.64 & 9+13 & 60& 18\%  +10\% -5\% & 48\% $\pm$6\% \\
\noalign{\smallskip}
\hline 
\end{tabular}

\label{samples}
\tablefoottext{a}{Planets hosts from HARPS (including the Sun)+planet hosts from other surveys.}\\
\tablefoottext{b}{There are 2 more stars in this \teff\ range with metallicities -1.04 and -1.07 shown in Table \ref{tabla_harps_comp}.}\\

\end{table*}      
\end{center}

\section{Observations and stellar samples}

The principal sample used in this work is formed by 1111 FGK stars observed within the context of the HARPS GTO programs to search for planets. The stars in that project were selected from a volume-limited stellar sample observed by the CORALIE spectrograph at La Silla observatory \citep{udry00}. The stars were selected to be suitable for radial velocity surveys. They are slowly-rotating and non-evolved solar type dwarfs with spectral type between F2 and M0 which also do not show high level of chromospheric activity.  The final sample is a combination of three HARPS sub-samples hereafter called HARPS-1 \citep{mayor03}, HARPS-2 \citep{locurto} and HARPS-4 \citep{santos_harps4}. Note that the HARPS-2 planet search program is the complementation of the previously started CORALIE survey \citep{udry00} to fainter magnitudes and to a larger volume.\\

The individual spectra of each star were reduced using the HARPS pipeline and then combined with IRAF\footnote{IRAF is distributed by National Optical Astronomy
Observatories, operated by the Association of Universities for
Research in Astronomy, Inc., under contract with the National
Science Fundation, USA.} after correcting for its radial velocity. The final spectra have a resolution of R $\sim$115000 and high signal-to-noise ratio (55\% of the spectra have S/N higher than 200), depending on the amount and quality of the original spectra. The total sample is composed by 135 stars with planets and 976 stars without detected planets but in our effective temperature region of interest (5600-5900 K) we have 42 and 235 stars with and without planets. 
To increase the number of stars with planets we have used spectroscopic data for 49 planet hosts which come from different observing runs, listed in Table~\ref{tabinstr}, some of them belonging to the CORALIE survey (see Table \ref{tab_otros}). The data reduction was made with the IRAF package or with the respective telescopes pipelines. All the images were flat-field corrected, sky substracted and added to obtain 1D spectra. Doppler correction was also done. We note that our sample contains 98\% of planets hosts within the range 5600-5900 K discovered by radial velocities and 77\% if we also consider the transiting planet hosts \citep{santos13}.

\begin{figure}
\centering
\includegraphics[width=9.0cm]{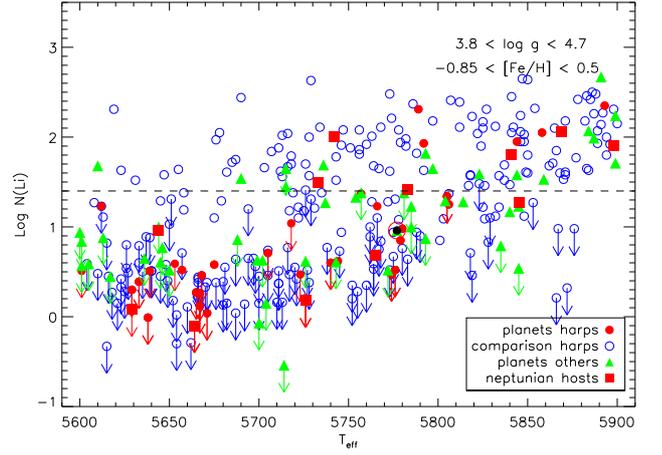}
\caption{Lithium abundances vs. \teff ~for planet host stars (red filled circles) and "single" stars (blue open circles) from HARPS together with other planet hosts (green triangles). Squares are stars which only host Neptunian- or Super-Earth-type planets. Down arrows represent A(Li) upper limits. The straight line at A(Li) = 1.4 matches the upper envelope of the lower limits for most of our stars.} 
\label{Liteff_todas}
\end{figure}

\section{Analysis}
      
The stellar atmospheric parameters were taken from \citet{sousa08,sousa_harps4,sousa_harps2} for HARPS stars and from \citet{santos04, santos05, sousa06,mortier13_transits} for the rest of the planet hosts. The errors of the parameters from \citet{santos04, santos05} are of the order of 44 K for $T_{\rm eff}$, 0.11 dex for $\log g$, 0.08 km s$^{\rm -1}$ for $\xi_{t}$, 0.06 dex for metallicity and 0.05M$_{\odot}$ for the masses. From \citet{sousa08,sousa_harps4,sousa_harps2} the typical errors are 30 K for $T_{\rm eff}$, 0.06 dex for $\log g$, 0.08 km s$^{\rm -1}$ for $\xi_{t}$ and 0.03 dex for metallicity. We refer to those works for further details in the parameters determination and errors. All the sets of parameters were determined in a consistent way as to reduce at maximum the systematic errors. We note here the uniformity of the adopted stellar parameters as discussed in Section 5 of \citet{sousa08}. Our group is continuously updating the stellar parameters of stars under consideration \citep{sousa08,sousa_harps4,sousa_harps2} and has created an online catalogue \citep{santos13} in order to guarantee a homogeneous and precise study of stellar abundances. This is especially important for Li when we have clearly limited the study to solar analogs.\\

Li abundances, ${\rm A(Li)}$\footnote{${\rm A(Li)}=\log[N({\rm Li})/N({\rm H})]+12$}, were derived by a standard LTE analysis using spectral synthesis with the revised version of the spectral synthesis code MOOG2010 \citep{sneden} and a grid of Kurucz ATLAS9 atmospheres with overshooting \citep{kurucz}.
We used the linelist from \citet{ghezzi_li6} though we applied a slight correction to the log \textit{gf} (to -2.278) of FeI line at 6707.4 \AA{} to adjust the Kurucz Solar Flux Atlas spectrum \citep{kurucz84}. We neglected possible $^{6}$Li contributions. We did not apply NLTE corrections since for this kind of stars they are quantitatively insignificant compared to the large dispersion of Li abundances or to a conservative typical error of 0.1 dex \citep{takeda05,ramirez_li12}. Some examples of spectral synthesis are shown in Fig. \ref{Li_synthesis}.
All abundances and their errors are listed in Tables~\ref{tabla_harps_plan}, \ref{tabla_harps_comp}\footnote{We note that there are some stars in common between the three used HARPS samples. We only used the best available spectrum for each of those, with its corresponding parameters as shown in these tables.} and \ref{tab_otros} .\\

Stellar masses and ages were derived by using the stellar evolutionary models from the Padova group computed with the web interface\footnote{http://stev.oapd.inaf.it/cgi-bin/param} dealing with stellar isochrones and their derivatives to the stars of our sample. The values for \teff\ and [Fe/H] were taken from the previously mentioned works and V magnitudes and parallaxes come from the Hipparcos database. For those stars with no values in Hipparcos database we used other sources as Simbad database\footnote{http://simbad.u-strasbg.fr/simbad/sim-fid} or The Extrasolar Planets Encyclopaedia\footnote{http://exoplanet.eu/catalog/} \citep{schneider}.


\section{Discussion}

In Fig.~\ref{Liteff_todas} we present a general overview of the behaviour of Li as a function of effective temperature for solar type stars. We can see in the plot the ranges in \teff\~, [Fe/H] and gravity for the stars in this sample (see also Table \ref{samples}).
As expected, below 5650K, most of the stars have severe depleted Li abundances regardless of whether they have planets or not. These stars have deeper convective envelopes that allow the material to reach the Li burning layers on the MS. However, a similar depletion of Li is also observed in stars with slightly higher temperatures (around solar \teff). Standard models \citep{deliyannis90,pinsonneault97}, which only consider mixing by convection, do not predict the Li depletion observed in solar-type stars. Indeed, the location of the base of the solar convection zone inferred from helioseismology is not hot enough to burn Li on the MS \citep{christensen91}.\\

On the other hand, when the temperature is close to 5900 K, most of the stars have preserved higher amounts of Li (due to their shallower convective envelopes) but still present signs of depletion that are not expected from canonical models either \citep[e.g.][]{pinsonneault97}. Therefore, there must be other mechanisms like rotationally induced-mixing \citep{pinsonneault90}, diffusion \citep{richer93}, internal waves \citep{montalban00}, overshooting \citep[e.g.][]{xiong09} or the combination of several \citep[e.g.][]{chaboyer95,charbonnel05} which account for the destruction of Li observed in solar-type stars in open clusters and in the field \citep[e.g.][]{pace12}.\\


\begin{figure}
\centering
\includegraphics[width=9.0cm]{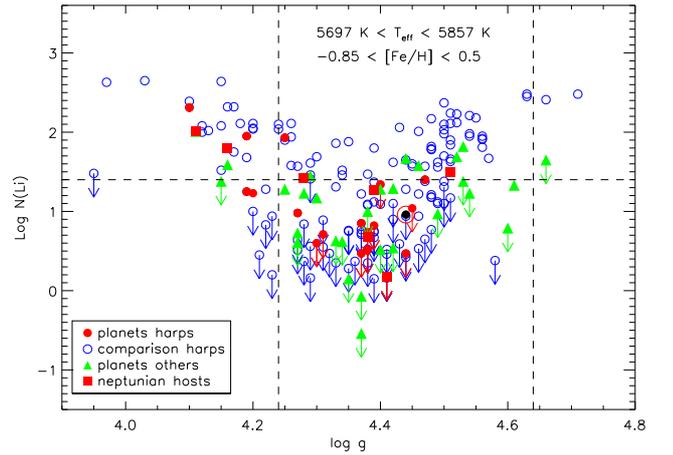}
\caption{Lithium abundances vs. \logg\ for solar type stars. Symbols as in Fig. \ref{Liteff_todas}. The vertical lines denote the range in \logg\ for our solar analogues and solar twins.} 
\label{Lilogg}
\end{figure}

\begin{figure}
\centering
\includegraphics[width=9.0cm]{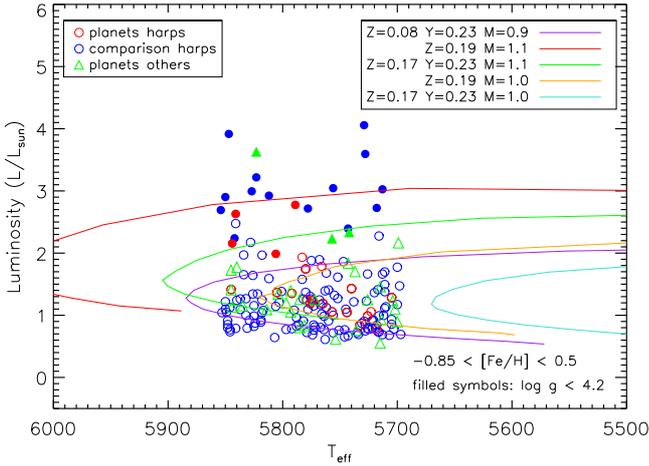}
\caption{HR diagram for solar-type stars. Evolutionary tracks are from \cite{girardi,bertelli}. Filled symbols are stars with \logg\ values lower than 4.2.} 
\label{luminosidad}
\end{figure}

As seen in Fig.~\ref{Liteff_todas}, there is a high dispersion in Li abundances in the \teff\ interval 5600-5900 K. Most of the planet hosts have destroyed their Li in this \teff\ window. 
To better appreciate the effects of planets on Li depletion we focus now in the solar temperature range, \teff\ = T$_{\odot} \pm$ 80 K, where previous works found differences between both groups of stars \citep[e.g.][]{israelian09}. In this range of temperatures the convective envelope is not as deep as in cooler stars but lies very close to the Li burning layer, hence, if some mechanism exists capable of producing an extra mixing (even if it is not very intense), Li will suffer a certain depletion. As a consequence, this type of stars are very sensitive to non-standard mixing processes.\\ 

Below we will discuss the dependence of Li depletion on several parameters. 
It is important to note that to make a meaningful comparison of planet hosts and "single" stars we have to deal with solar-type stars in the MS. Therefore, we removed from our sample of planet hosts some stars showing activity and/or with very young ages like HD 81040, which has a disc \citep{sozzetti06}; HD 70573 \citep{setiawan07} or Corot-2 \citep{alonso08} as well as subgiants or slightly evolved stars like HD 118203 and HD 149143 \citep{dasilva06}; HD38529 \citep{fischer01}; HD 48265 \citep{minniti09}; HD 175167 \citep{arriagada10} ; HD 219828 \citep{melo07}. Some of these stars have been included by other authors \citep{gonzalez_li10,ghezzi_li,ramirez_li12} in previous works about Li depletion and planets, however, we prefer to discard them since they are in different evolutionary statuses. For instance, a subgiant with solar \teff\ had a higher temperature when it was on the MS and thus different Li content.\\

We want to stress that our comparison sample of "single" stars is formed by stars that have been surveyed during years (with some of the most precise planet search projects) and no planets have been detected so far. We are confident that most of those stars do not host nearby giant planets. 
We cannot rule out that most of "single" stars in our sample host low mass planets that are below detection capabilities of HARPS. It is also possible that some of these stars have long-period giant planets yet to be discovered.
This is an important point since some earlier studies \citep{ryan00,chen06,luck06,gonzalez_li10,baumann,ramirez_li12} were using field stars as "single" stars, not included in planet-search surveys, but they could have giant planets and thus are not the best targets to be used as real comparison stars.
We also point that by keeping the comparison sample from HARPS and adding more planet hosts from other surveys, our results will be statistically more significant. Nevertheless we will differentiate those stars in the plots to show that we still keep the homogeneity of the study.

\begin{figure*}
\centering
\includegraphics[width=13.0cm]{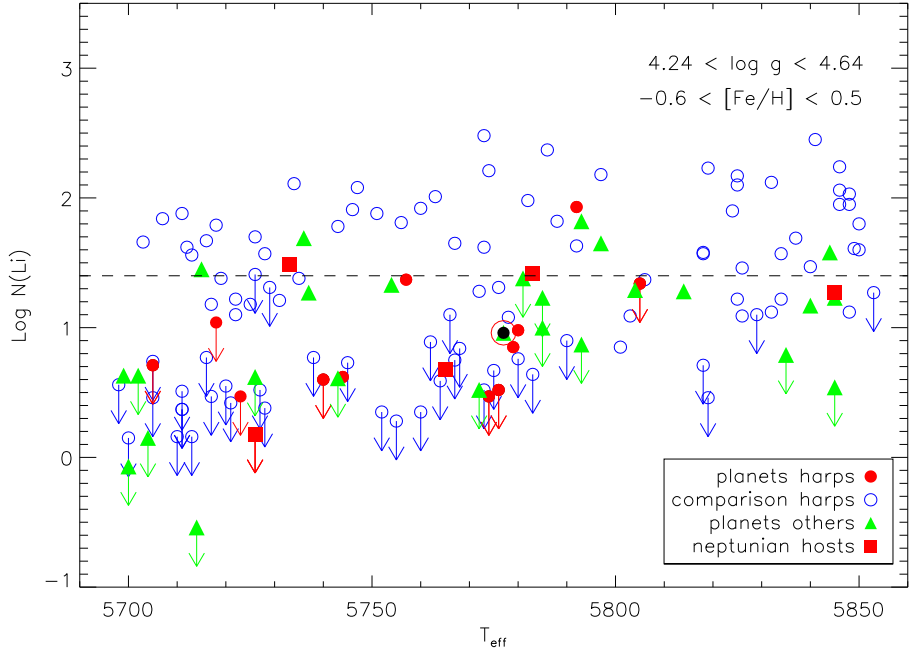}
\caption{Lithium abundances vs. \teff ~ for solar analogues). Symbols as in Fig. \ref{Liteff_todas}.} 
\label{Liteff_solar}
\end{figure*}

\subsection{Li and log g: cleaning the sample from evolved stars}
Our sample is mainly composed by G dwarfs with \logg\ values between 4.1 and 4.6 (Fig. \ref{Lilogg}). There are several stars with \logg\ values between 4.1 and 4.2 with high Li abundances, opposite to expected by the general trend starting at 4.2 dex. This was already noticed in \citet{baumann}. Indeed, all the giant planet hosts with high Li have low \logg\ values except HD 9446 (\teff\ = 5793 K, A (Li) = 1.82) and HD 181720 (\teff\ = 5792 K, A (Li) = 1.93). This last planet host has \logg\ = 4.25 but $L=2.65 L_{\odot}$ and an age of 11.2 Gyr, thus it is possible that this star is entering the subgiant phase. The general trend in Fig. \ref{Lilogg} shows a decrease in Li abundances with decreasing \logg\, in the range 4.2 $<$ \logg\ $<$ 4.7, since these stars with lower \logg\ are older and have had more time to deplete their Li. If we split this plot in different metallicity bins we still observe this effect.\\

To check the evolutionary status of those stars we plot them in a HR diagram together with some evolutionary tracks from \citet{bertelli,girardi}. The luminosity was computed by considering the Hipparcos parallaxes, V magnitude and the bolometric correction as in \citet{sousa08}. In Fig. \ref{luminosidad} we can see that those stars with lower \logg\ values (filled symbols) have much higher luminosities on average and thus may be slightly evolved from the MS. The higher Li abundances of these stars could be explained by the fact that they were hotter when they were on the MS and hence did not deplete so much Li. Moreover, there are claims \citep{deliyannis90} that, such early evolution may bring up Li to the photosphere in subgiant stars from a reservoir (hidden below the base of the convective envelope) that could be formed due to microscopic diffusion during the lifetime in the MS. Therefore, to avoid these possibly evolved stars, we will restrict our study to stars with higher \logg\ values, and more specifically to stars with \logg\ = \logg$_{\odot}$\footnote{\logg$_{\odot}$ = 4.44 dex} $\pm$ 0.2, as showed by the vertical lines in Fig. \ref{Lilogg}. We note that 81\% of our stars with \teff\ = T$_{\odot} \pm$ 80 K have also \logg\ = \logg$_{\odot}$ $\pm$ 0.2. We will call this subsample 'solar analogues' (see Table \ref{samples}).

\begin{figure}
\centering
\includegraphics[width=9.0cm]{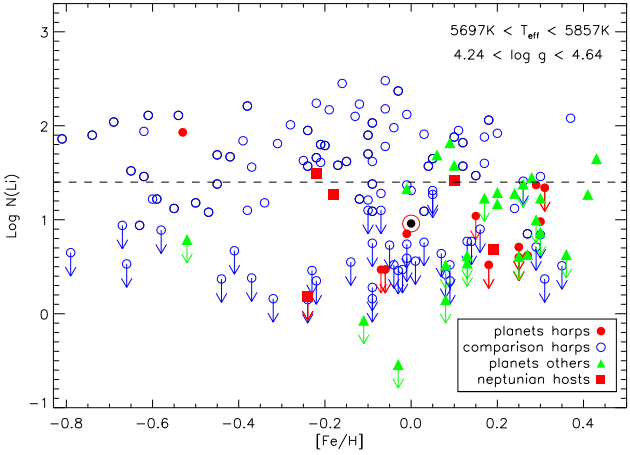}
\caption{Lithium abundances vs. [Fe/H] for solar analogues. Symbols as in Fig. \ref{Liteff_todas}.} 
\label{Lifeh_solar}
\end{figure}

\subsection{Li and \teff}
In Fig. \ref{Liteff_solar} we plot the final sample of solar analogues (see Table \ref{samples}). We have made a cutoff in [Fe/H]=-0.6 because our most metal-poor planet host has a metallicity of -0.53. In any case, only 10\% of our "single" stars within solar \teff\ and \logg\ range have [Fe/H] $<$ -0.6. This sample of solar analogues is composed of 43 planet hosts (including the Sun), from which only 8 (i.e. $<$ 19\%) have Li abundance higher than 1.4. These stars have ages from 2 to 11 Gyr and metallicities from -0.53 to 0.43. All of them except HD 9446 have planets with masses lower than Jupiter's (see Section 4.7).\\

On the other hand, there are 99 stars without known giant planets, from which 47\% have Li abundances higher than 1.4. Therefore, for this \teff\ range, there is a clear lack of planet hosts with high Li content. This fact is not caused by a difference in \teff\ , hence, the dependence on other parameters will be analyzed later.
We stress that given the homogeneous nature of the HARPS sample (and CORALIE survey) a priori there is no reason to expect that planet hosts should have lower Li in this domain (see Section 4.5). 'Single' stars are homogeneously distributed in this plot but planet hosts are not, although all of them have similar temperatures, gravities, metallicities and are on the MS.

\begin{figure*}
\centering
\includegraphics[width=9.0cm]{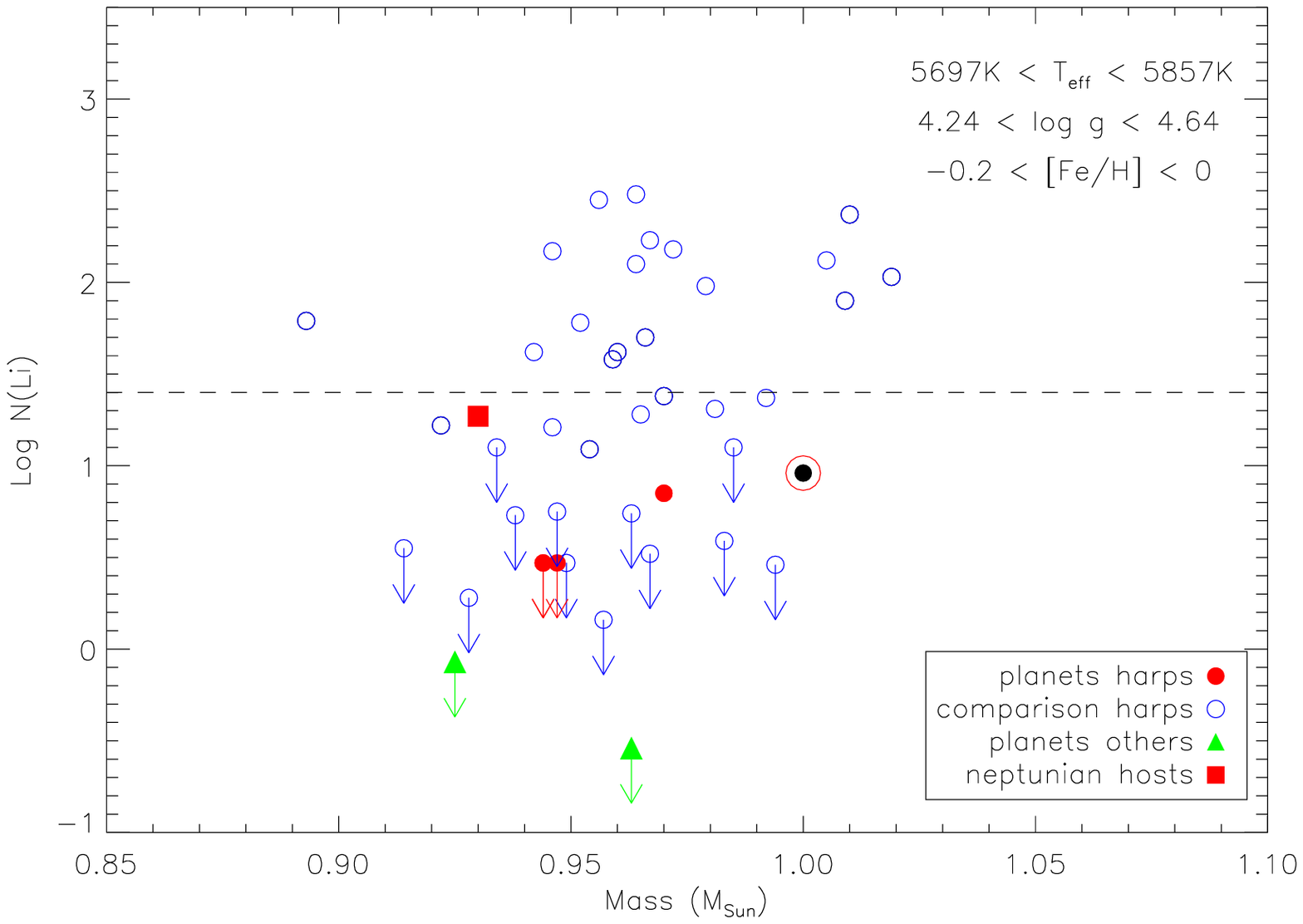}
\includegraphics[width=9.0cm]{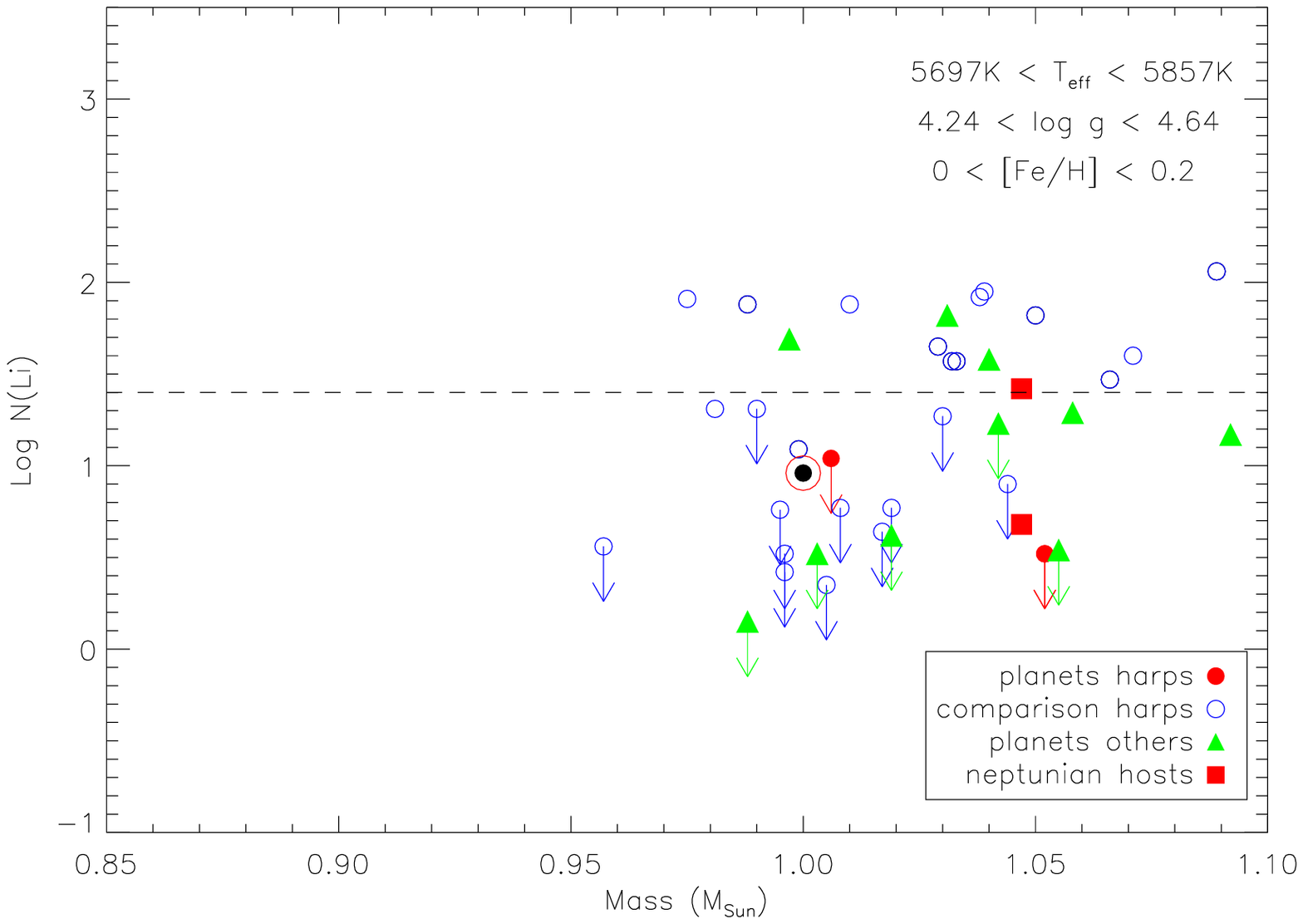}
\caption{Lithium abundances vs. mass for solar twins at [Fe/H] $<$ 0 (left panel) and solar twins at [Fe/H] $>$ 0 (right panel). Symbols like in Fig. \ref{Liteff_todas}.}
\label{Limass_solar}
\end{figure*}

\subsection{Li and [Fe/H]}

In Fig. \ref{Lifeh_solar} we show Li abundances as a function of metallicity for what we defined above as solar analogues. Most of the planet hosts present high [Fe/H] values although a few of them reach values lower than -0.5. The increase of metal opacities in solar-type stars is responsible for the transition between radiative and convective energy transport. Thus, stars with more metals (at a given mass) are expected to have deeper convective envelopes which favour Li depletion though this assumption might not be correct \citep{pinsonneault01}. Therefore, we could first think that the only reason of planet hosts to be Li depleted is because they are metal rich. This might be true for very metal rich stars. We can see in Fig. \ref{Lifeh_solar} how above 0.2 dex both stars with and without planets deplete a lot of lithium though we still can detect Li absorption in some stars. This fact suggests that in our solar \teff\ range, high metallicity is also playing a role in lithium depletion which could be comparable to the role of planets, hence we cannot extract any conclusion from the most metallic stars. However, we can see in the plot that at lower metallicities most of the planet hosts also present low Li abundances while "single" stars are homogeneously spread along high and low Li abundances. Indeed, if we focus on solar twins (metallicities within 0.2 dex the solar value, see Table \ref{samples}) only 4 out of 22 planet hosts show Li higher than 1.4 while 29 out of 60 "single" stars have A (Li) $>$ 1.4. We can also observe that the [Fe/H] distribution at low and high Li abundances of both groups is quite homogeneous.  We do not find a trend of Li with metallicity in this region (-0.2 $<$ [Fe/H] $<$ 0.2). Therefore, we can be sure that for solar twins [Fe/H] is not playing a major role in the depletion of Li.\\

\begin{figure*}
\centering
\includegraphics[width=9.0cm]{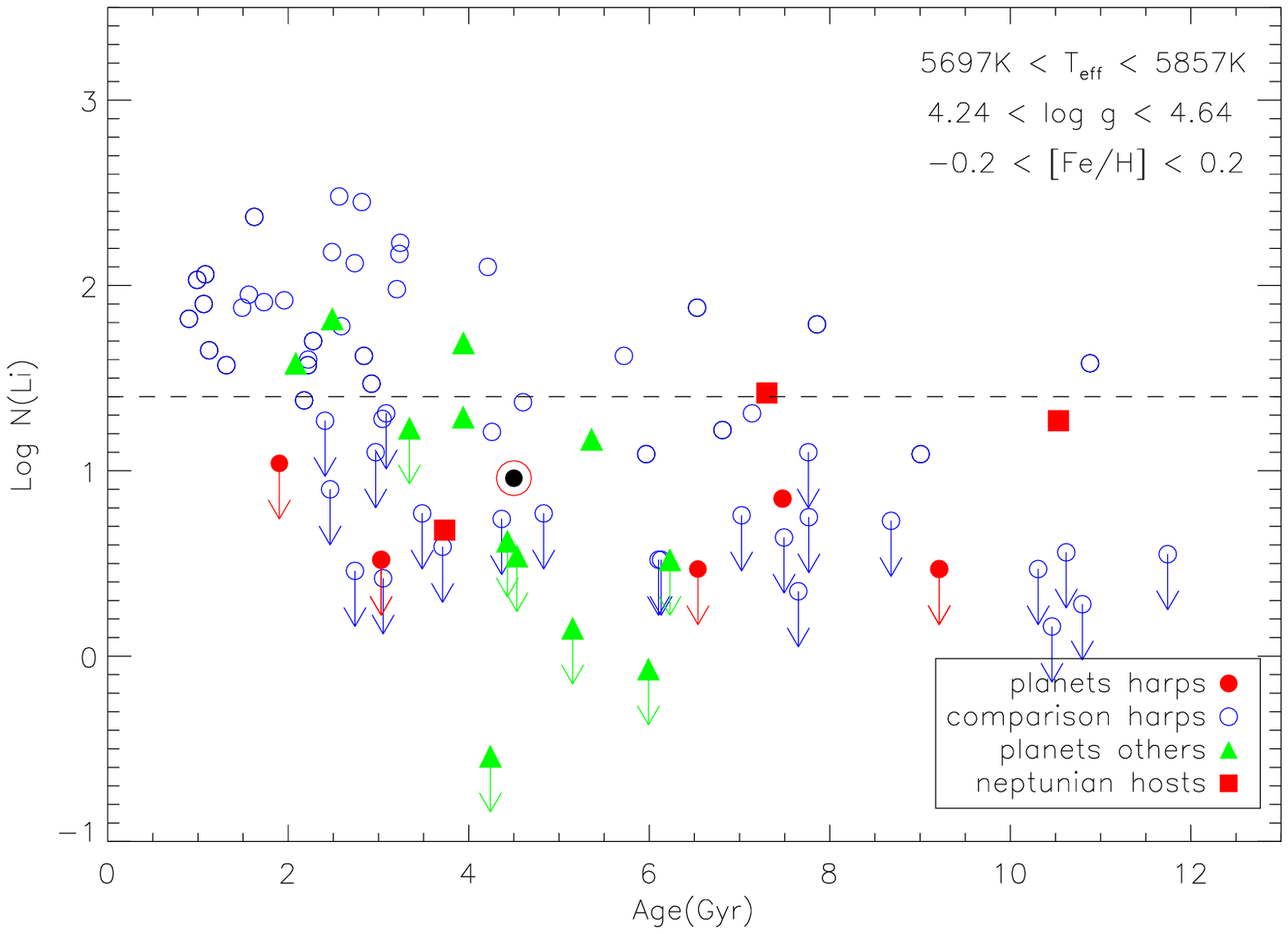}
\includegraphics[width=9.0cm]{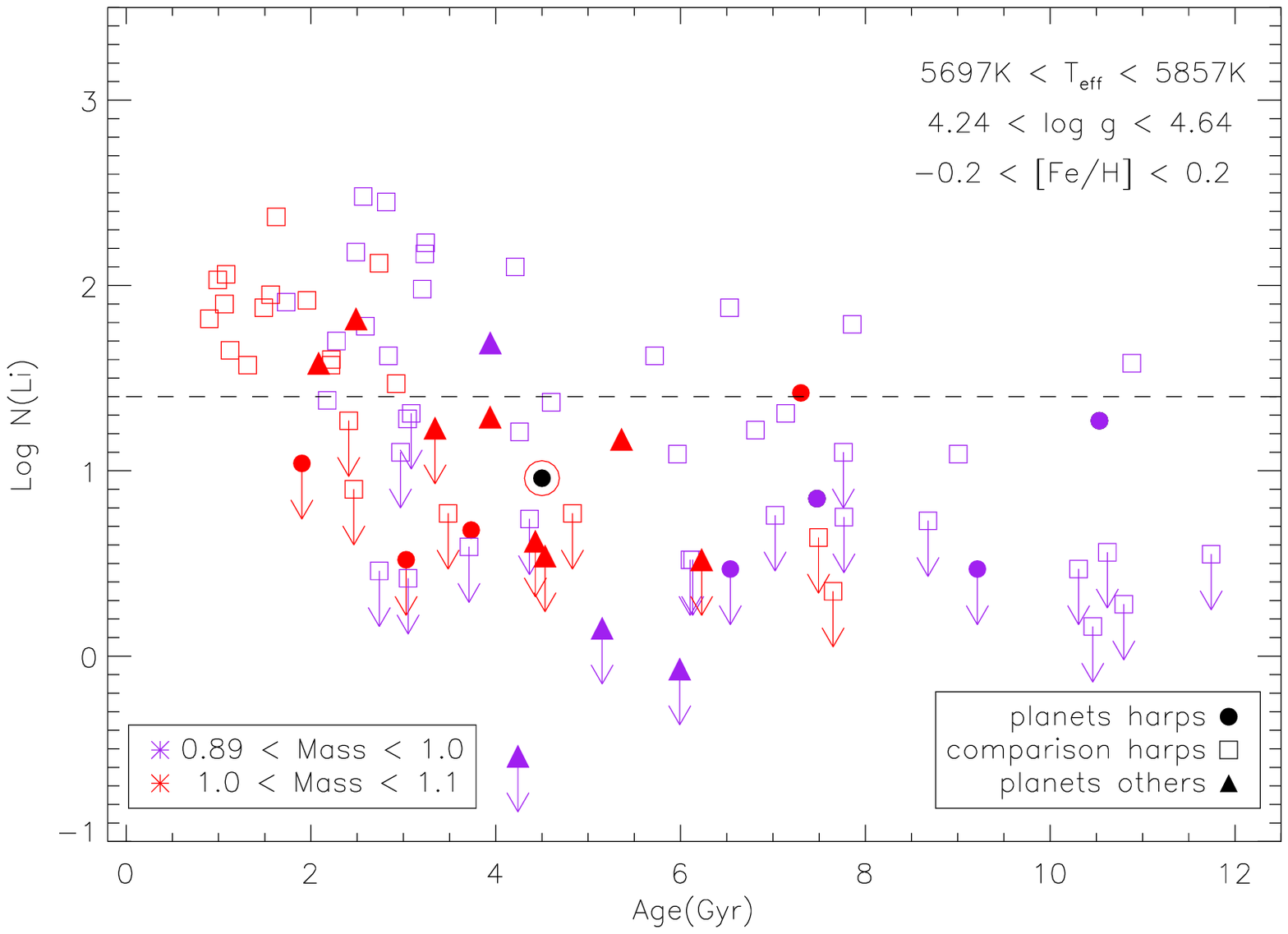}
\caption{Lithium abundances vs. age for solar twins (left panel) and for solar twins with colours denoting different mass ranges (right panel).}
\label{Liage_solar}
\end{figure*}

\subsection{Li and stellar mass}

On the Main Sequence, stellar mass is directly correlated with \teff\ (with a dependence on [Fe/H]), i.e. mass increases with \teff, although solar type stars show a wide spread in masses. However, it is not correct to use only mass as a parameter to constrain solar analogues because some stars with masses and [Fe/H] close to solar have high temperatures (up to 6000 K) and thus fall out of our solar range. We prefer to limit our sample by their \teff\ since it is a parameter directly observed and its determination has lower uncertainties than the determination of mass, which in turn depends on the uncertainties of the stellar parameters and metallicity and lies on the theoretical evolutionary tracks.\\

We have seen in the previous section that in order to discard a metallicity effect on Li abundances we have to exclude metal rich stars ([Fe/H] $>$ 0.2) from our sample. On the other hand, at [Fe/H] $<$ -0.2 there are only 4 planet-hosts so, in the following, we will focus in what we defined solar twins (see Table \ref{samples}).
In Fig. \ref{Limass_solar} we present the Li relation with mass for solar twins stars in two metallicity regions. In general, the stars are homogeneously spread at several masses regardless of their Li depletion at both metallicity ranges though the average mass of planet hosts is higher due to their higher average [Fe/H]. We can also observe that stars with similar masses in each metallicity region present a high dispersion in Li abundandes. All the planet hosts with higher Li abundances are situated in the metal-rich panel (0 $<$ [Fe/H] $<$ 0.2)  but the average abundance of Li detections, 1.33 dex, is lower than for the "single" stars, 1.69. Even if we remove the younger stars ($<$ 1.5 Gyr), "single" stars have on average 0.30 dex higher abundances (1.33 vs 1.63, see section 4.5 and Table \ref{averages} for further discussion about Li dependence on age). For the lower metal-poor region (-0.2 $<$ [Fe/H] $<$ 0)  the difference is more obvious. Here we find that the average of Li detections for planet hosts is 1.02 whereas for "single" stars it is 1.79. If we discard young stars ($<$ 1.5 Gyr) the difference is practically the same. This result does not depend on stellar mass. We do not find any trend of Li abundances with this parameter and the average masses of both samples are very similar (see Table \ref{averages}).\\

This spread has been noticed in previous works, like for instance, \citet{pace12}, where they show how Li abundances do not depend on mass for solar type stars (one solar mass or lower) in M67 and they suggest that an extra variable, in addition to mass, age and metallicity, is responsible for the scatter in Li abundances, since those stars are very similar. A similar spread in Li is observed for field stars around one solar mass and -0.2 $<$ [Fe/H] $<$ 0 in \citet{lambert04}. Looking at these plots it is very clear that planet hosts have destroyed more Li regardless of their \teff, mass or metallicity, hence there must exist another parameter which explains this trend.

\begin{center}
\begin{table*}
\caption{Mean values of parameters, together with standard error of the mean, for different subsamples, all of them with \teff\ = T$_{\odot} \pm$ 80 K, and \logg\ = \logg$_{\odot}$ $\pm$ 0.2. Metallicity and age ranges (which include the boundaries) are indicated in the table. PH and CS denote planet hosts (from HARPS and other spectrographs) and comparison sample stars, respectively.}
\centering
\begin{tabular}{crccccccccc}
\hline
\hline
\noalign{\smallskip}
Group & Number & [Fe/H] range & Age range & \teff\ & [Fe/H] & \logg\ & Mass & Age & Li detections \\
 &  & Gyr & (K)  &   &     (cm\,s$^{-2}$) & M$_{\odot}$ & Gyr & A(Li)  \\
\noalign{\smallskip}
\hline    
\noalign{\smallskip}
A & 10 PH & [-0.2,0.2] & [1.5,12] & 5797$\pm$12 & 0.07$\pm$0.04 & 4.41$\pm$0.03 & 1.02$\pm$0.01 & 5.14$\pm$0.82 & 1.27$\pm$0.12  \\
A & 30 CS & [-0.2,0.2] & [1.5,12] & 5783$\pm$8 & -0.04$\pm$0.02 & 4.47$\pm$0.02 & 0.98$\pm$0.01 & 4.01$\pm$0.44 & 1.75$\pm$0.07  \\
\noalign{\smallskip}
\hline    
\noalign{\smallskip}
B &  5 PH & [-0.2,0.2] & [4,12] & 5805$\pm$15 &  0.02$\pm$0.06 & 4.36$\pm$0.03 & 1.01$\pm$0.03 & 7.03$\pm$1.04 & 1.13$\pm$0.10  \\
B &  11 CS & [-0.2,0.2] & [4,12] & 5768$\pm$15 & -0.07$\pm$0.03 & 4.40$\pm$0.02 & 0.96$\pm$0.01 & 6.63$\pm$0.62 & 1.48$\pm$0.10  \\
\noalign{\smallskip}
\hline    
\hline    
\noalign{\smallskip}
C &   3 PH & [-0.2,0] & [1.5,12] & 5800$\pm$22 & -0.06$\pm$0.06 & 4.40$\pm$0.02 & 0.97$\pm$0.02 & 7.50$\pm$1.74 & 1.03$\pm$0.12  \\
C &  22 CS & [-0.2,0] & [1.5,12] & 5778$\pm$9  & -0.09$\pm$0.01 & 4.47$\pm$0.02 & 0.96$\pm$0.01 & 4.33$\pm$0.53 & 1.77$\pm$0.09  \\
\noalign{\smallskip}
\hline    
\noalign{\smallskip}
D &   3 PH & [-0.2,0] & [4,12] &  5800 $\pm$22 & -0.06$\pm$0.06 & 4.40$\pm$0.02 & 0.97$\pm$0.02  & 7.50$\pm$1.74   & 1.03$\pm$0.12  \\
D &   9 CS & [-0.2,0] & [4,12] &  5770 $\pm$17 & -0.10$\pm$0.02 & 4.41$\pm$0.03 & 0.95$\pm$0.01  & 6.72$\pm$0.76   & 1.48$\pm$0.11  \\
\noalign{\smallskip}
\hline
\hline     
\noalign{\smallskip}
E &  8 PH & [0,0.2] & [1.5,12] & 5793$\pm$13 & 0.12$\pm$0.03 & 4.42$\pm$0.03 &  1.04$\pm$0.01 & 4.17$\pm$0.58 & 1.33$\pm$0.13  \\
E &  9 CS & [0,0.2] & [1.5,12] & 5795$\pm$16 & 0.09$\pm$0.02 & 4.45$\pm$0.03 &  1.02$\pm$0.01 & 3.58$\pm$0.10 & 1.63$\pm$0.10  \\
\noalign{\smallskip}
\hline
\hline     
\noalign{\smallskip}
F &  3 PH & [0,0.2] & [4,12] & 5800$\pm$20 & 0.10$\pm$0.06 & 4.34$\pm$0.05 &  1.05$\pm$0.02 & 5.72$\pm$0.83 & 1.18$\pm$0.13  \\
F &  3 CS & [0,0.2] & [4,12] & 5763$\pm$27 & 0.04$\pm$0.03 & 4.36$\pm$0.02 &  0.99$\pm$0.01 & 6.54$\pm$0.34 & 1.43$\pm$0.24  \\
\noalign{\smallskip}
\hline    
\hline    
\noalign{\smallskip}
G &   8 PH & [-0.25,0.15] & [1.5,12] & 5786$\pm$15 & -0.01$\pm$0.04 & 4.44$\pm$0.03 & 0.99$\pm$0.02 & 5.85$\pm$1.07 & 1.38$\pm$0.12  \\
G &  35 CS & [-0.25,0.15] & [1.5,12] & 5789$\pm$8  & -0.09$\pm$0.02 & 4.46$\pm$0.02 & 0.96$\pm$0.01 & 4.52$\pm$0.50 & 1.75$\pm$0.07  \\
\noalign{\smallskip}
\hline    
\noalign{\smallskip}
H &   5 PH & [-0.25,0.15] & [4,12] & 5783$\pm$18 & -0.06$\pm$0.06 & 4.40$\pm$0.04 & 0.97$\pm$0.03 & 7.65$\pm$0.98 & 1.20$\pm$0.13  \\
H &  16 CS & [-0.25,0.15] & [4,12] & 5779$\pm$14 & -0.12$\pm$0.03 & 4.40$\pm$0.02 & 0.95$\pm$0.01 & 6.87$\pm$0.57 & 1.53$\pm$0.07  \\
\noalign{\smallskip}
\hline    
\noalign{\smallskip}
I &   4 PH  & [-0.25,0.15] & [5,12] & 5785$\pm$23 & 4.39$\pm$0.05 & -0.08$\pm$0.08  & 8.45$\pm$0.74 & 0.96$\pm$0.03 &     1.26$\pm$0.14 \\
I &   12 CS & [-0.25,0.15] & [5,12] & 5777$\pm$17 & 4.38$\pm$0.02 & -0.12$\pm$0.03  & 7.73$\pm$0.57 & 0.94$\pm$0.01 &     1.52$\pm$0.08 \\
\noalign{\smallskip}
\hline    
\noalign{\smallskip}

\end{tabular}

\label{averages}
\end{table*}      
\end{center}

\subsection{Li and age}
Some previous works have claimed that the observed Li depletion in planet hosts is a consequence of the evolution of the star \citep[e.g.][]{baumann} and thus it depends on age. Therefore, if planet hosts were systematically older than single stars that might be the reason for their lower Li abundances. We want to emphasize that for stars with ages between ~1.5 and 8 Gyr the width of spectral lines do not change considerably since v \textit{sin} i is nearly constant during MS, \citep[e.g.][]{pace04}. Therefore, for radial velocity surveys these are all good targets for precise planet searches regardless of their age. 
We agree that Li is destroyed as the star gets older but this depletion takes place principally during the first 1 Gyr \citep{randich10} (if we consider logarithmic abundances) and depends on initial rotation rates \citep[e.g.][]{charbonnel05} whereas after 1-2 Gyr the age effect is not so strong. For instance, the models by \citet{deliyannis97}, which include rotational induced mixing, show Li abundances of $\sim$2.9 dex, $\sim$2.1 dex and $\sim$1.7 dex at ages of 100 Myr, 1.7 Gyr and 4 Gyr, respectively, for a star with \teff\ = 5800K and initial rotation v \textit{sin} i = 10 km/s. Furthermore, if age were the principal cause on Li depletion we would not observe such a dispersion ($\sim$1.3 dex) in Li abundances for stars in the same evolutionary status in clusters like M67, where all the stars have the same age and metallicity \citep{pasquini08,randich09,pace12}. In the clusters NGC 3960 \citep{prisinzano}, Collinder 261 \citep{pallavicini} and NGC 6253 \citep{randich10} a Li dispersion in solar type stars was found as well, though not as large as in M67. On the other hand, the near solar metallicity cluster NGC188 of 6-8 Gyr present abundances 10-20 times higher than in the Sun \citep{randich03}.\\

In the left panel of Fig. \ref{Liage_solar} we explore the relation of Li with age for solar twins. We observe that there is not a tight relation with age except for very young ages. Furthermore, we can find young stars with low Li abundances as well as old stars with higher abundances. Both groups of stars are spread over the entire range of ages though at younger ages (0-2 Gyr) there are more "single" stars than planet hosts. This is likely due to the higher difficulties in finding planets around young, active stars. In order to test this correlation we computed a generalized Kendall's tau correlation coefficient considering upper limits of Li. This was done with the program R and the function \textit{cenken} which can be used with censored data \citep{akritas}. This test gives $\tau$ = 0.138 (with a probability of 0.63 of having a correlation) for planet hosts and $\tau$ = -0.004 for single stars (with P=0.03 of having a correlation). Therefore, this test shows that there is not a clear relationship between Li and age. However, if we deal only with the Li detections $\tau$ decreases to -0.24 for both groups, with a higher probability of correlation for "single" stars, 0.96 vs 0.63 for planet hosts. \\ 

In the right panel of Fig. \ref{Liage_solar} we can observe the same plot but split in two mass ranges, 0.89-1.0 M$_{\odot}$ and 1.0-1.1 M$_{\odot}$. In the more massive subsample (red symbols) we can only find detections of Li for "single" stars at ages $<$ 3 Gyr so we cannot compare stars at older ages. However, in that age bin (0-3 Gyr) none of the planet hosts can reach the average of Li for "single" stars. On the other hand, in the less massive subgroup (purple symbols) the number of stars is higher. Planet hosts are spread at different ages but in general present lower Li abundances than the "single" stars of similar ages. In any case, the stars of both mass subgroups might not be very different considering typical errors in mass of 0.1 M$_{\odot}$ \citep[e.g.][]{casagrande07,fernandes11}. In addition, we show in the previous section that for solar twins (0.9-1.1 M$_{\odot}$) Li does not depend on mass, even in narrower metallicity ranges, so mixing all those in the same plot (left panel of Fig. \ref{Liage_solar}) will not affect our conclusions.
Furthermore, we note that when dealing with Main-Sequence stars, the age determination is probably very uncertain \citep[e.g.][]{jorgensen2}, at least significantly more uncertain than the mass determination.\\ 

\begin{figure}
\centering
\includegraphics[width=9.0cm]{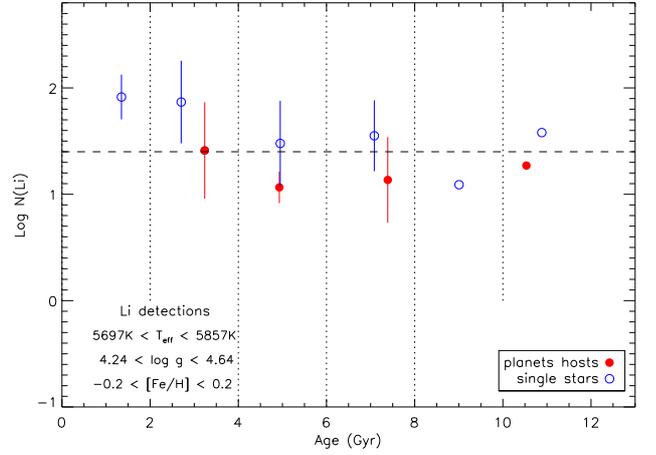}
\caption{Average of Li detections for solar twins in different age bins. The points in each bin are situated in their average age. The bars indicate the dispersion in Li abundances if there are more than one star in each bin.} 
\label{Liage_bins}
\end{figure}

Finally, in Fig. \ref{Liage_bins} we depict the average of Li detections for solar twins in different age bins. We can observe a decrease in Li abundances for solar age stars with respect to younger objects but stars with planets always present lower Li abundances in each age bin for ages $<$ 8 Gyr. At older ages we only have three objects so we do not contemplate those age bins.\\

In table \ref{averages} we sumarize the average values of parameters in several subsamples already discussed. We note that Li averages are calculated with only detections since for upper limits we cannot know the real content of Li. We discard the younger stars ($<$ 1.5 Gyr) to avoid a comparison sample too much biased towards low ages. For the solar twins (-0.2 $<$ [Fe/H] $<$ 0.2, group A in the table) the "single" stars present Li abundances nearly 0.5 dex higher than planet hosts and only the planet host HD 9446 present a Li abundance higher than the average for "single" stars. This difference in mean Li is higher than 3.4$\sigma$ (two side t-test) whereas the difference for age is 1.2$\sigma$. For older stars ($>$ 4 Gyr, group B) the difference is also high, 0.35 dex, but in this case none of the planet hosts reach A(Li) = 1.48. The planet host group has higher average [Fe/H] but that difference in metallicity is not enough to explain the difference in Li as already discussed in previous sections. Moreover, \citet{pinsonneault01} showed that the mass of the convective envelope (responsible for the degree of Li depletion) hardly varies for a wide range of metallicities at a given \teff\ . Nevertheless, we find different Li abundances if we split the solar twins in two metallicity regions. In the less metallic group (-0.2 $<$ [Fe/H] $<$ 0), ``single'' stars are very biased towards young ages (group C) so we must only take into account older stars (group D). For the more metallic stars (groups E and F) the mean ages and metallicities are more similar in the younger subsample (group E) and we find a difference of 0.3 dex in Li abundance.
Planet hosts are also slightly older on average ($\sim$ 1 Gyr) in most of the subsamples but the difference is at the level of the errors and, as we already pointed out, the main destruction of Li occurs before 1-2 Gyr. Since we are not considering those young objects the slight difference in average ages cannot justify the big differences in Li. We also look at a subsample of solar twins shifted to lower metallicities ([Fe/H] = -0.05 $\pm$ 0.2, groups G and H) and the difference in Li abundances is quite similar, 0.43 and 0.33 dex for the groups with ages $>$ 1.5 Gyr and $>$ 4 Gyr, respectively. In order to find more similar populations regarding age and metallicity we analyzed an older subsample: group I. The differences in average [Fe/H] and age are similar to group H, 0.04 dex and 0.7 Gyr, respectively, while the Li difference is decreased to 0.26 dex. This fact suggests that for older ages it is more difficult to detect differences though the number of stars in this subsample is significantly lower and thus the statistical inferences might be less reliable.\\

With the aim of increasing the number of stars in our sample we applied survival statistics to several subsamples from Table \ref{averages} which allows to extract information from the upper limits. We used the package ASURV \citep{feigelson} which provides the Kaplan-Meier estimator of the mean and several two-sample tests. For the group A, 4 out of 5 tests give probabilities lower than 0.06 that the samples of planets hosts and ``single'' stars come from the same parent population. The Kaplan-Meier estimator of the mean is 0.361 $\pm$ 0.208 for planets hosts and 1.069 $\pm$ 0.116 for stars without planets, hence it seems these 2 samples are really distinct. For the older solar twins (group B), however, all the two-sample tests give probabilities between 0.4-0.6. The K-M estimator of the mean for planet hosts is 0.104 $\pm$ 0.229 and for stars without detected planets it is 0.722 $\pm$ 0.135. Therefore for this subgroup there is no clear difference. For the solar twins sample at lower [Fe/H] (group G) we find similar results as for group A, with probabilities around 0.08 that the samples are drawn from the same population. The K-M estimators of the mean are 0.330 $\pm$ 0.233 and 1.094 $\pm$ 0.109 for planet hosts and ``single'' stars, respectively, thus these subsamples appear to be different. For the older stars (group H) the two-sample test probabilities increase to 0.2 but the K-M estimators of the mean are still quite different (0.081 $\pm$ 0.226 vs 0.804 $\pm$ 0.123). For the group I the two-sample test probabilities increase to 0.55 (and the K-M estimators of the mean are more similar) suggesting that the possible effect of planets might be diluted for very old stars as pointed before.\\

These tests including upper limits and the values from detections shown in Table \ref{averages} indicate that there seems to be a different distribution of Li abundances for planet hosts. However, depending on the building of the subsamples and the restrictions applied on age and metallicity the results can be different, maybe due to the low number of stars in each subsample. Therefore, a bigger sample with more similar stars would be desirable to fully confirm this distinction.

\subsection{Testing the differences with observed spectra}

\begin{figure*}
\centering
\includegraphics[width=0.32\linewidth]{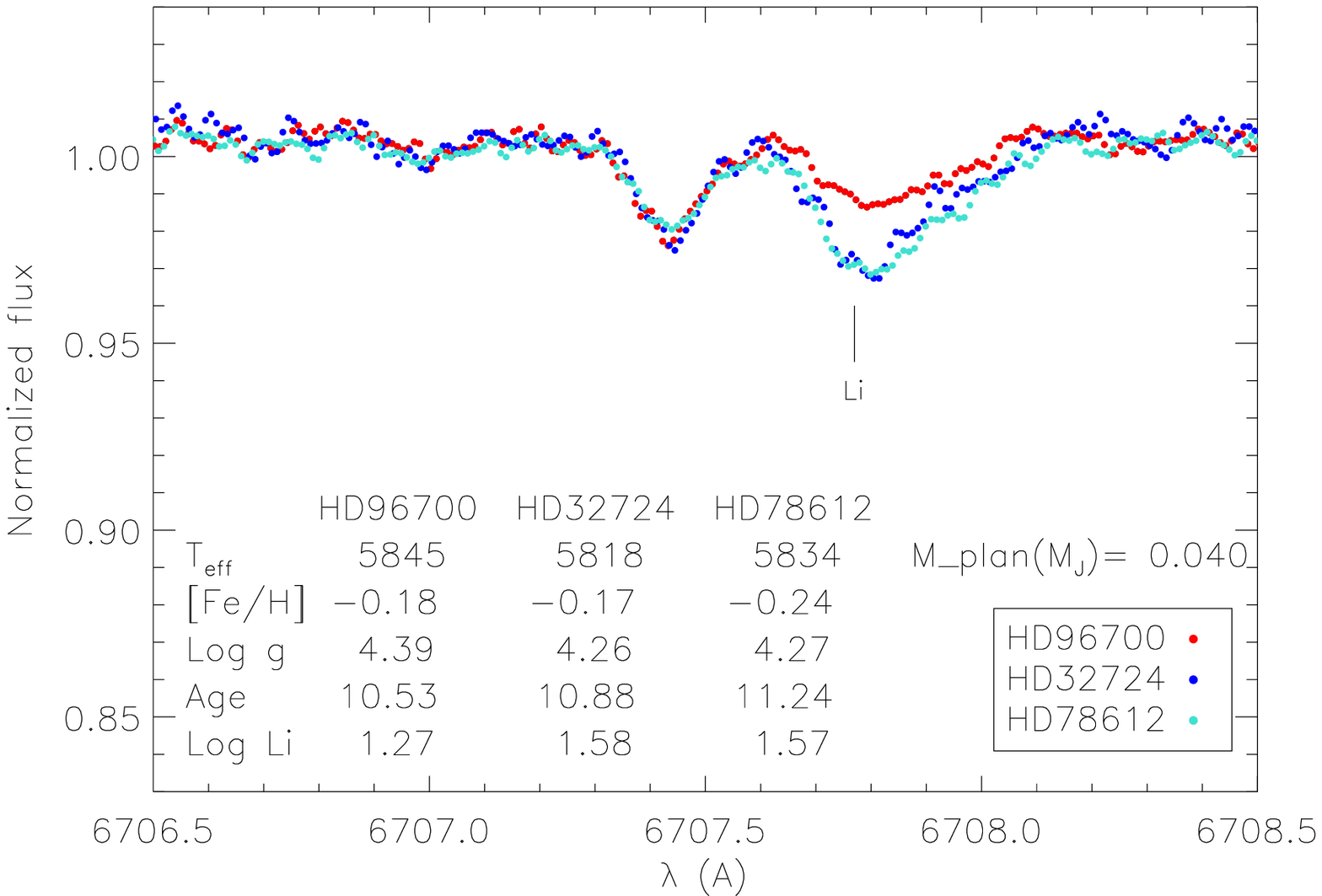}
\includegraphics[width=0.32\linewidth]{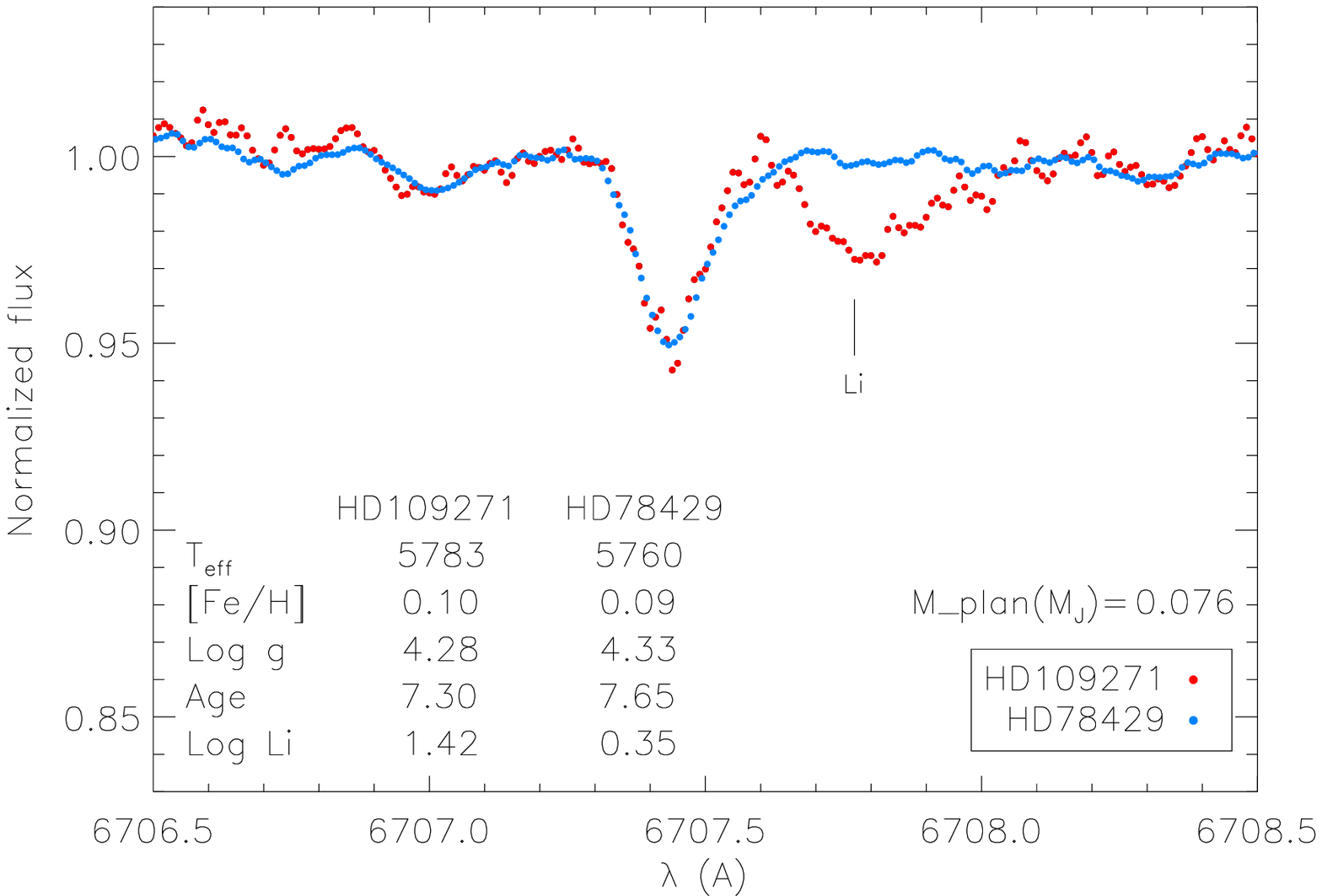}
\includegraphics[width=0.32\linewidth]{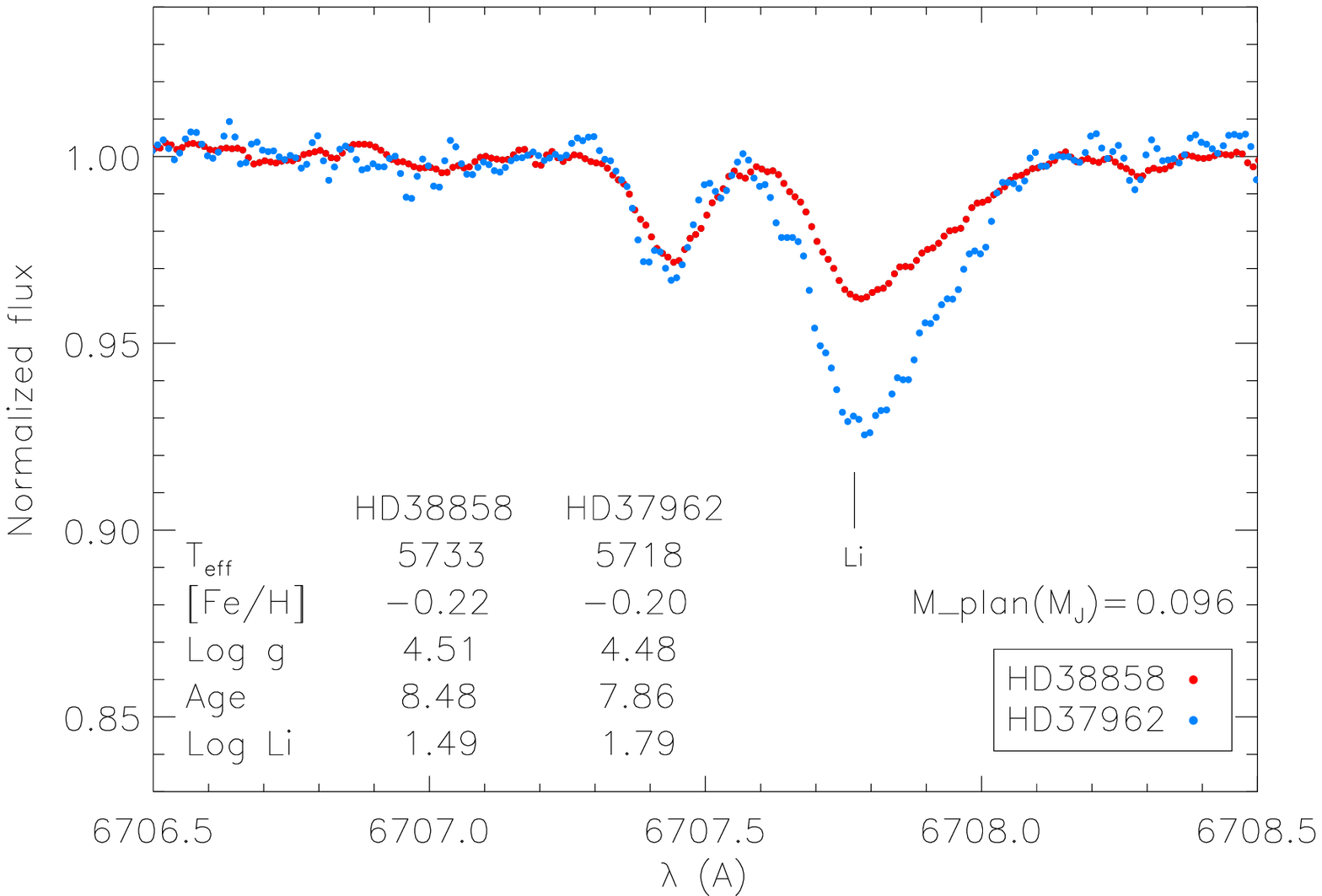}
\end{figure*}

\begin{figure*}
\includegraphics[width=0.32\linewidth]{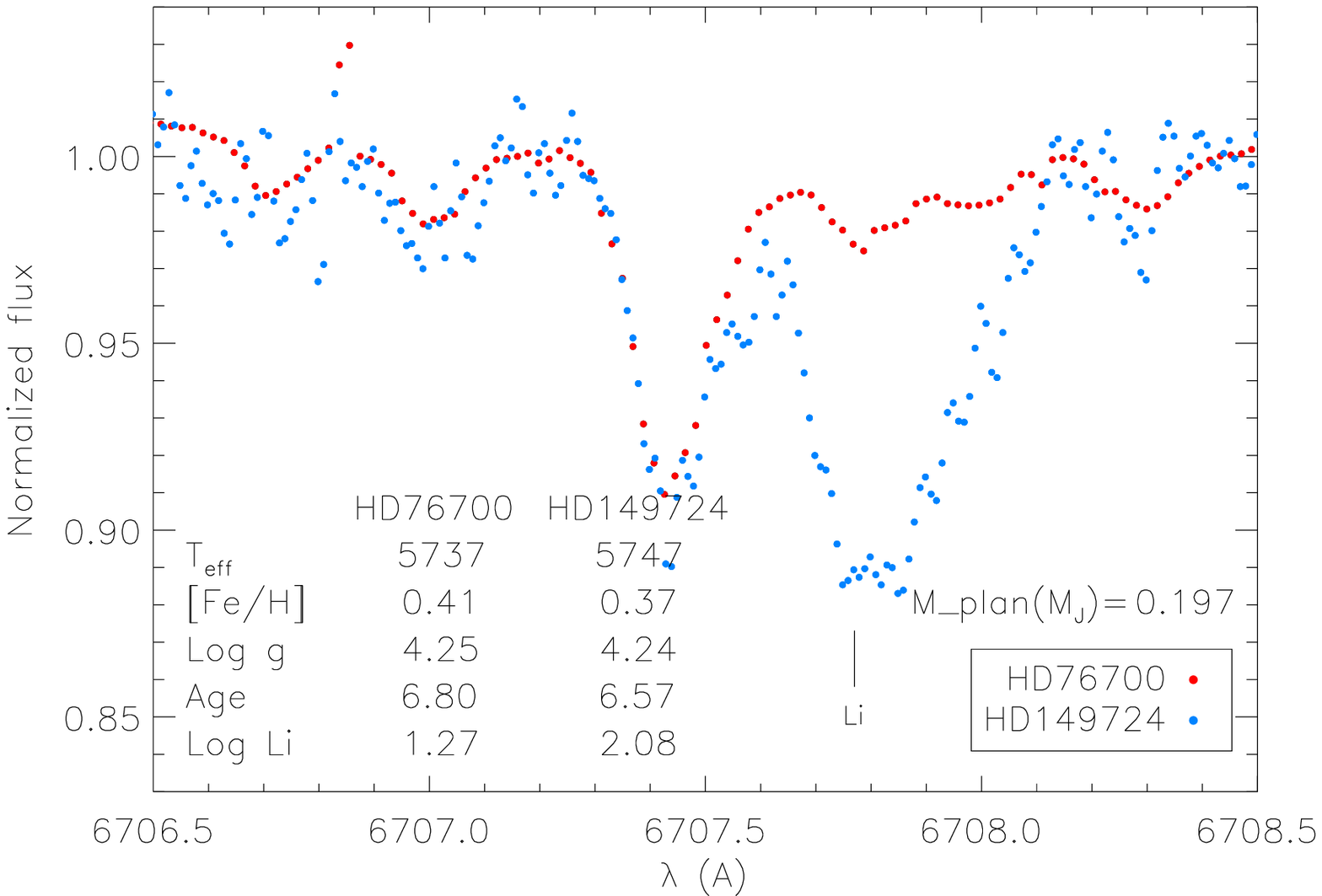}
\includegraphics[width=0.32\linewidth]{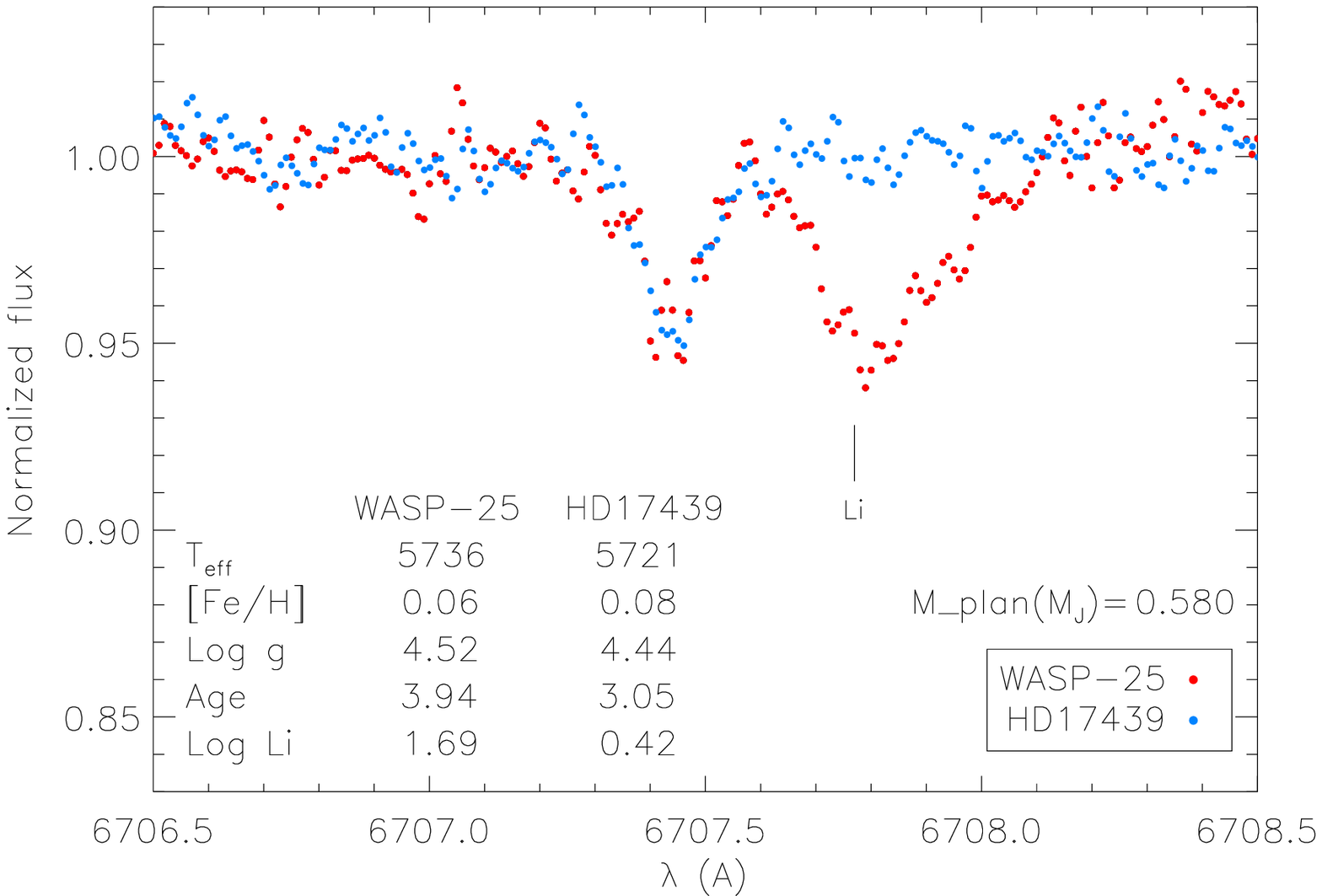}
\includegraphics[width=0.32\linewidth]{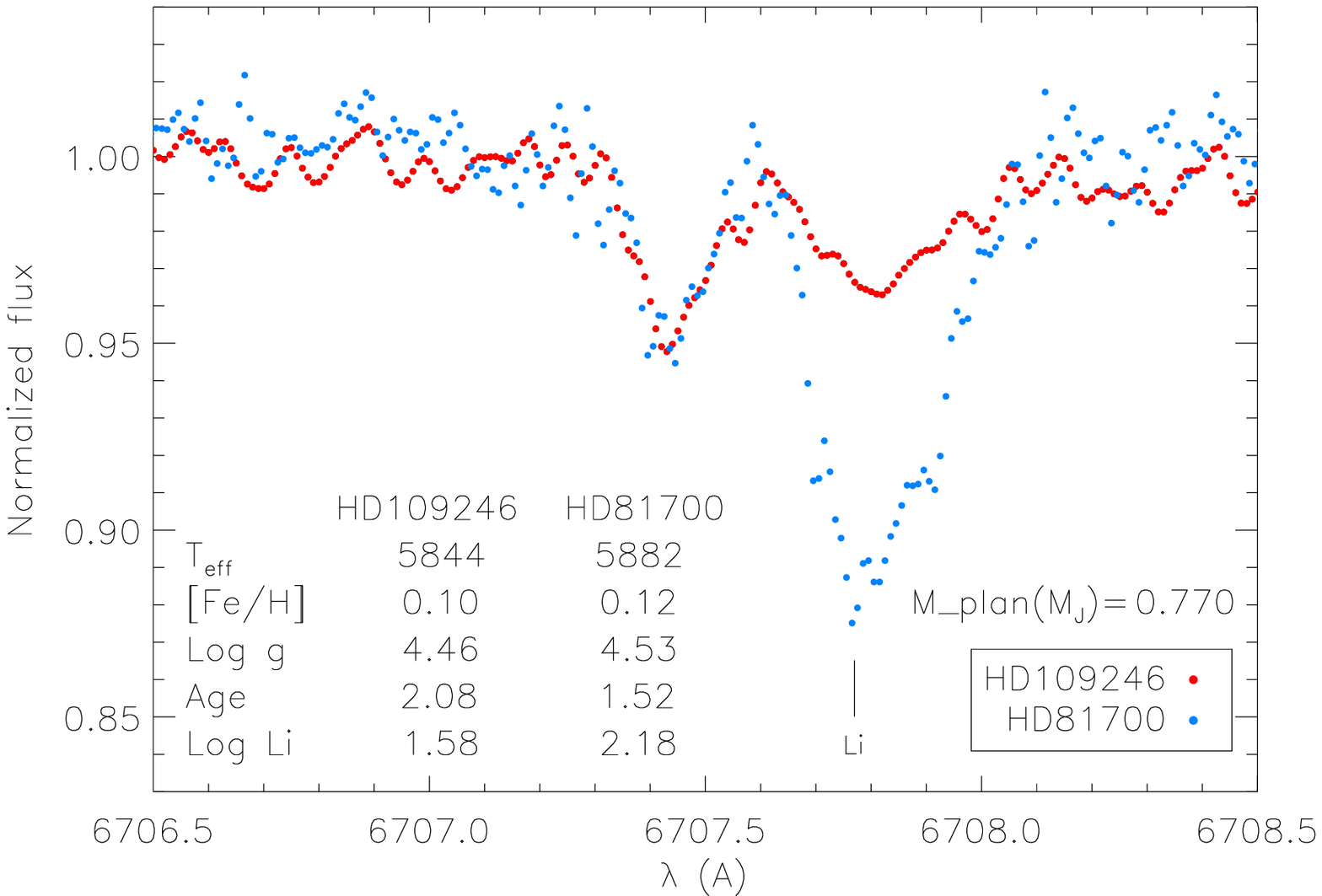}
\end{figure*}

\begin{figure*}
\includegraphics[width=0.32\linewidth]{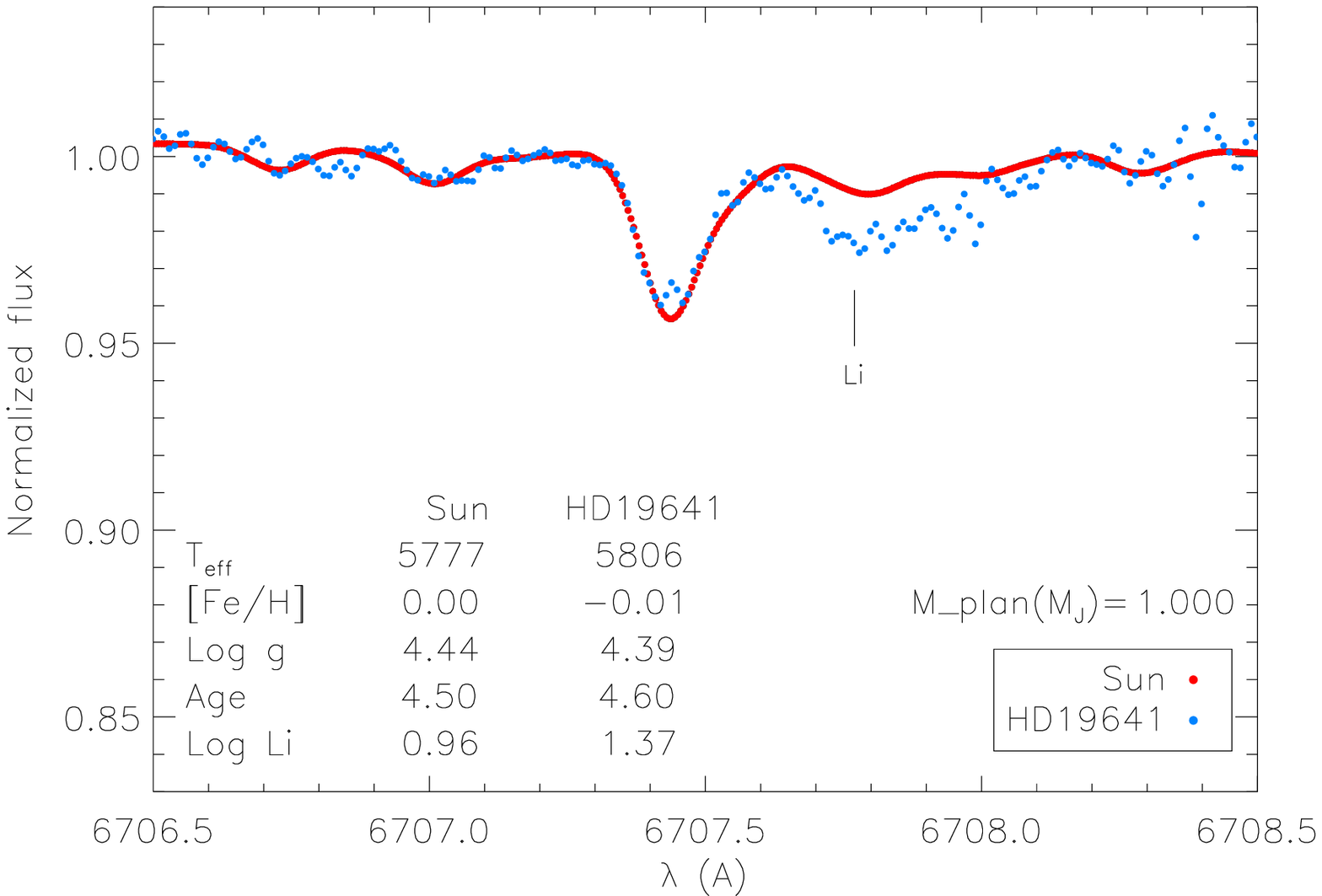}
\includegraphics[width=0.32\linewidth]{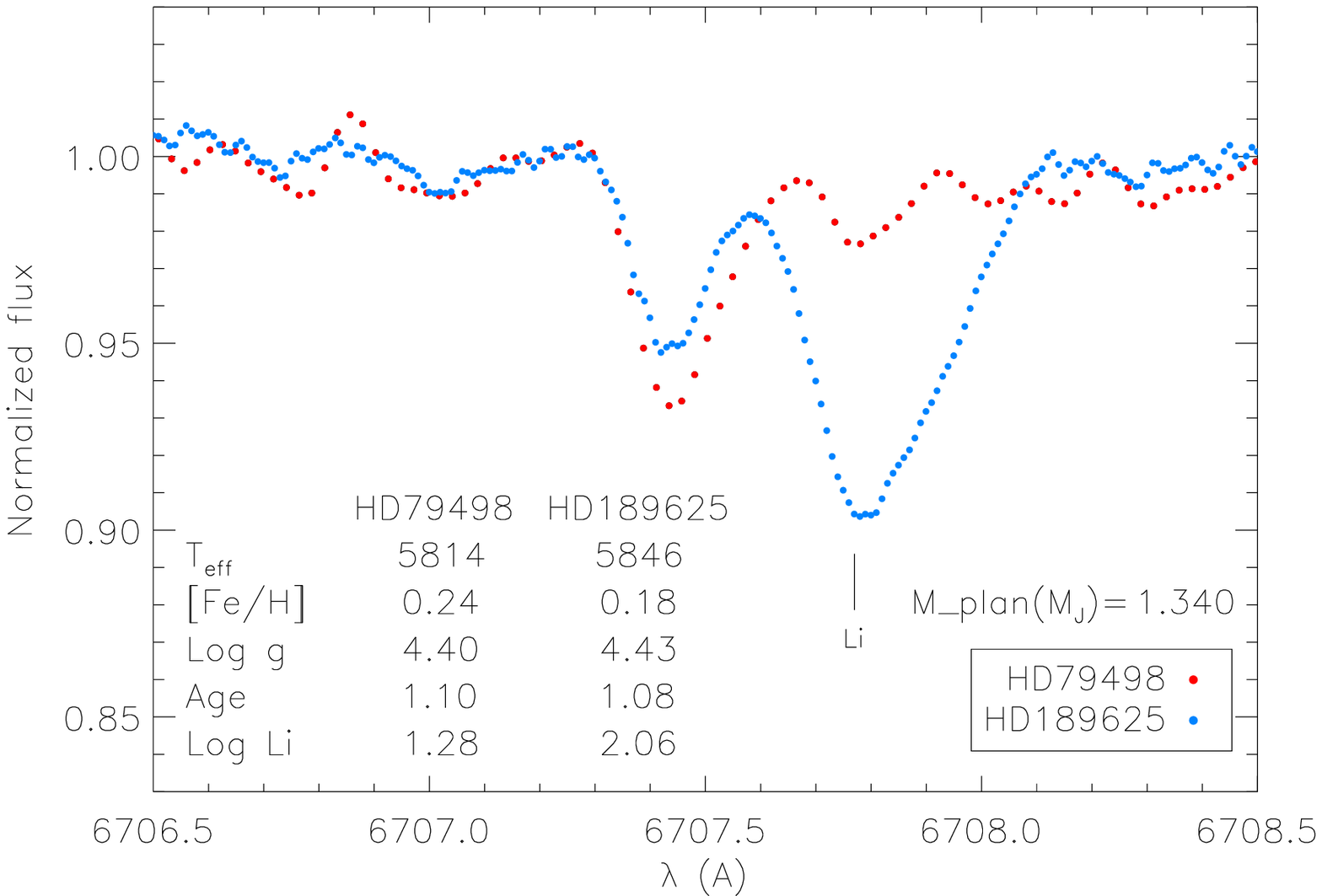}
\includegraphics[width=0.32\linewidth]{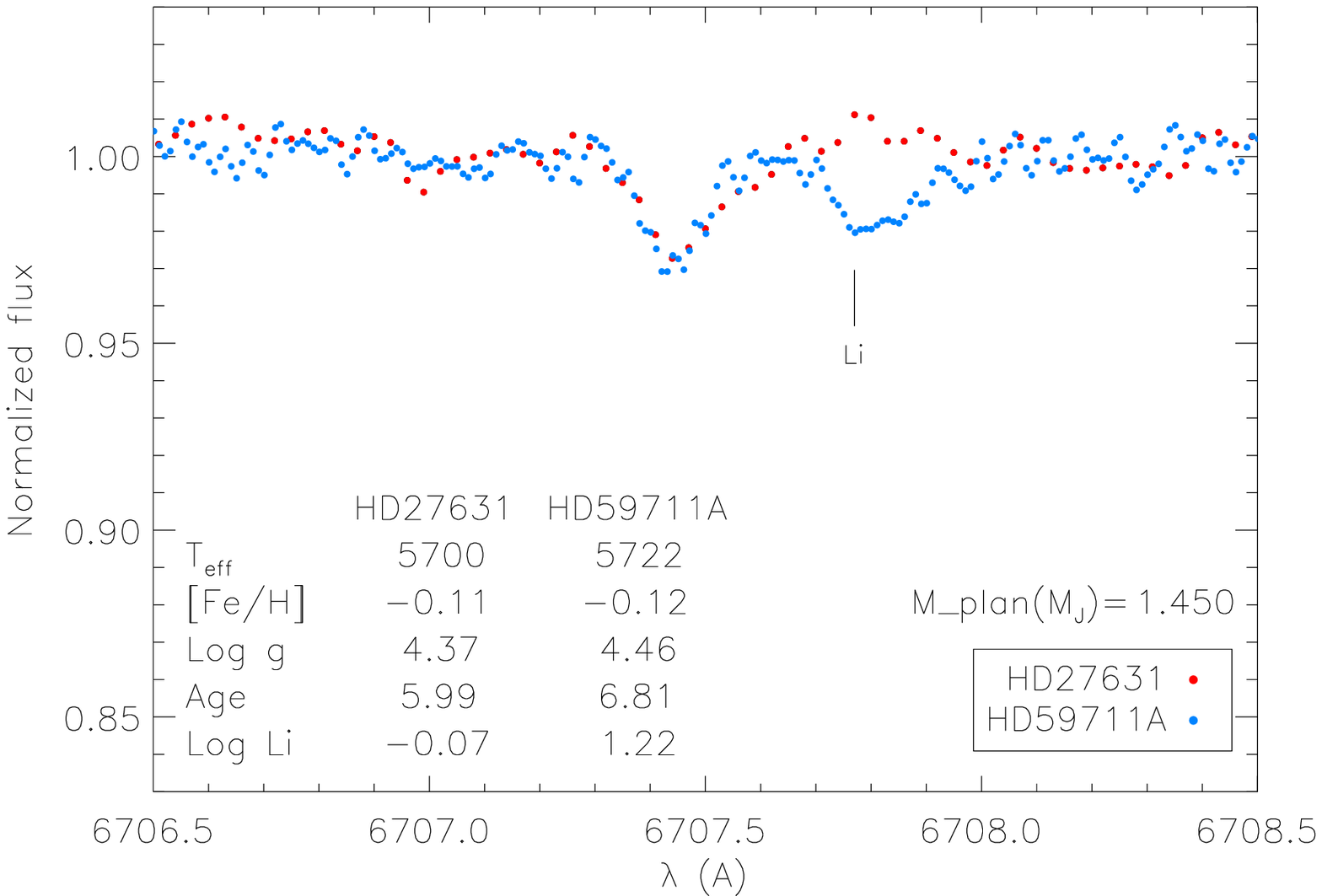}
\end{figure*}

\begin{figure*}
\centering
\includegraphics[width=0.32\linewidth]{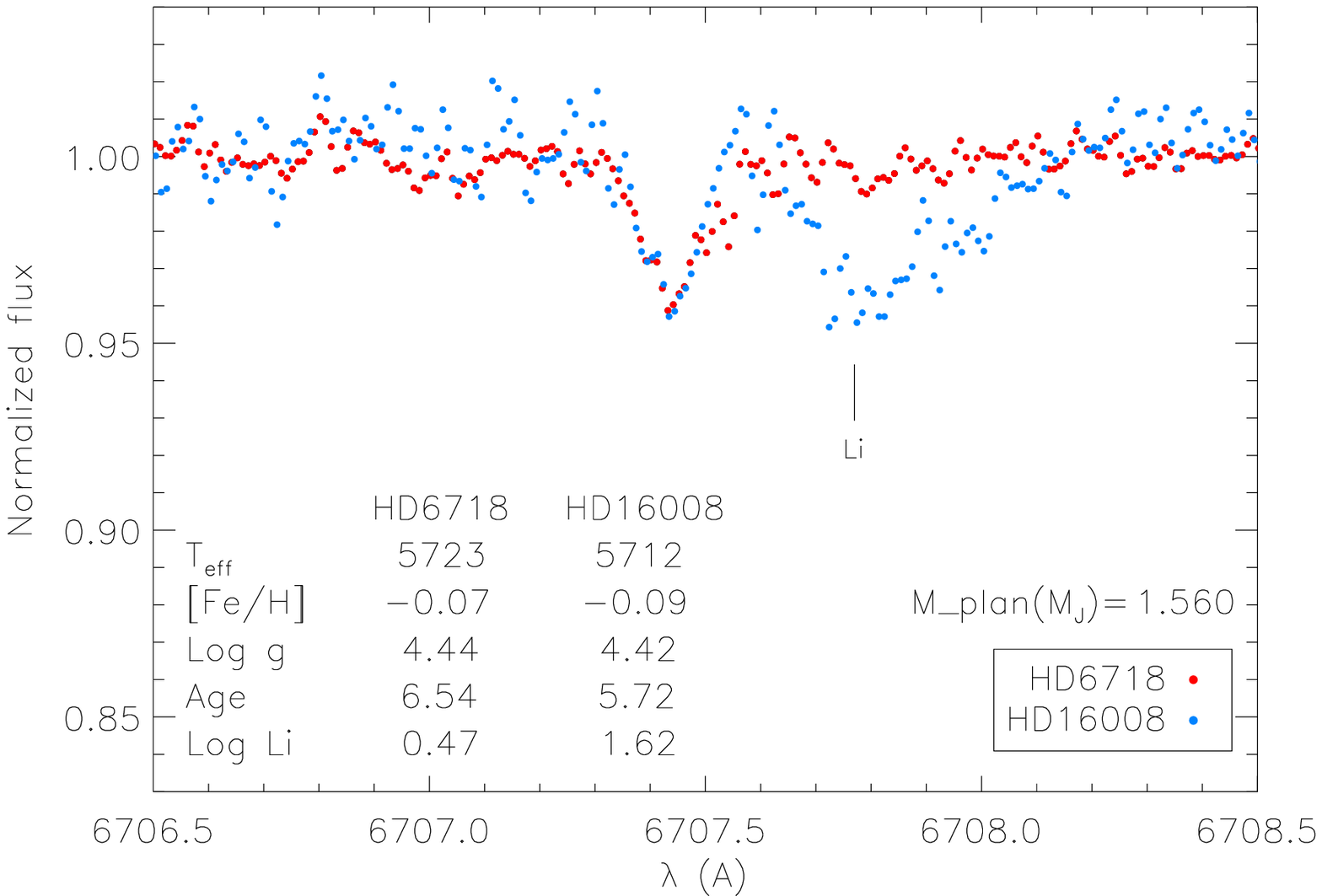}
\includegraphics[width=0.32\linewidth]{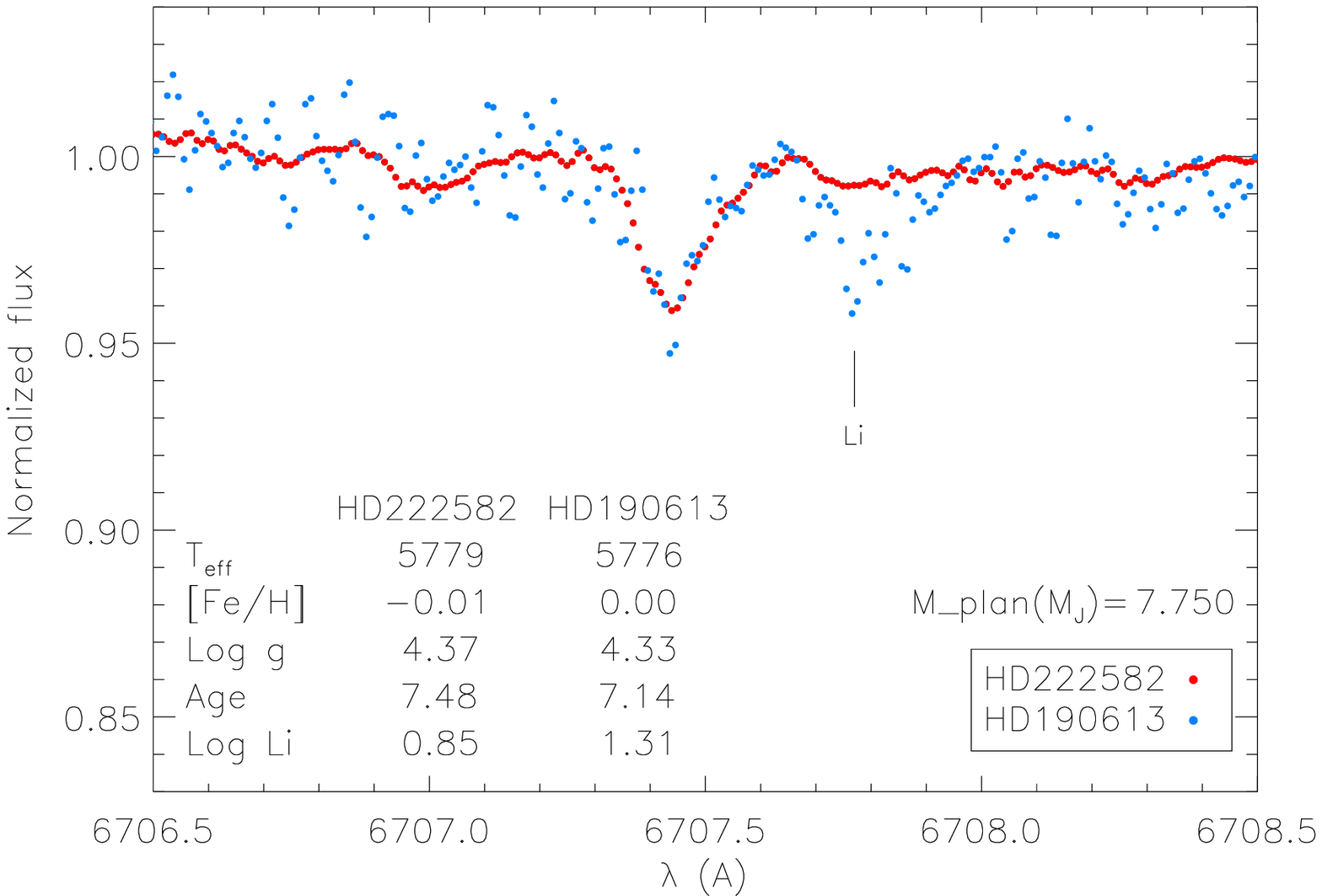}
\includegraphics[width=0.32\linewidth]{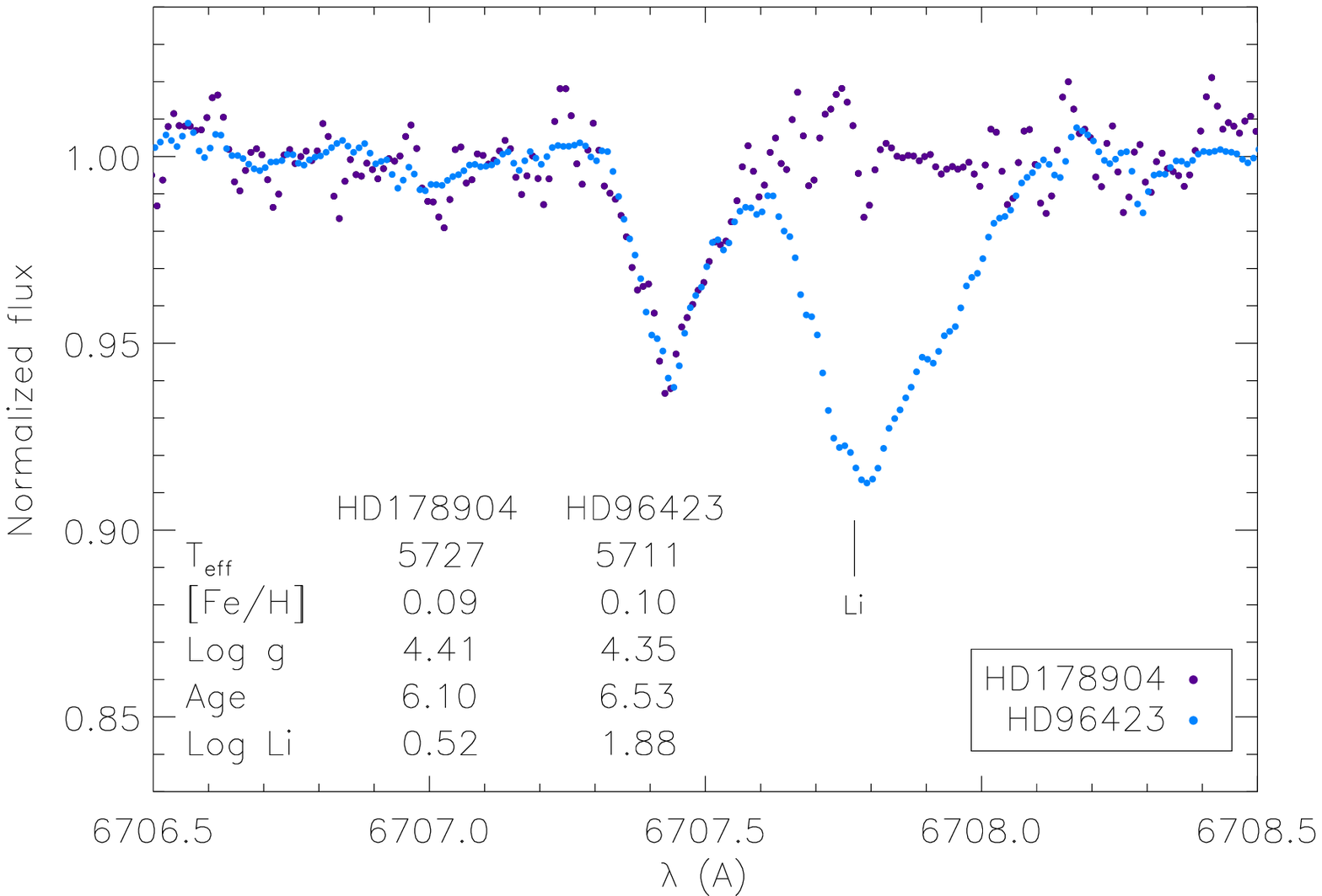}
\end{figure*}

\begin{figure*}
\centering
\includegraphics[width=0.32\linewidth]{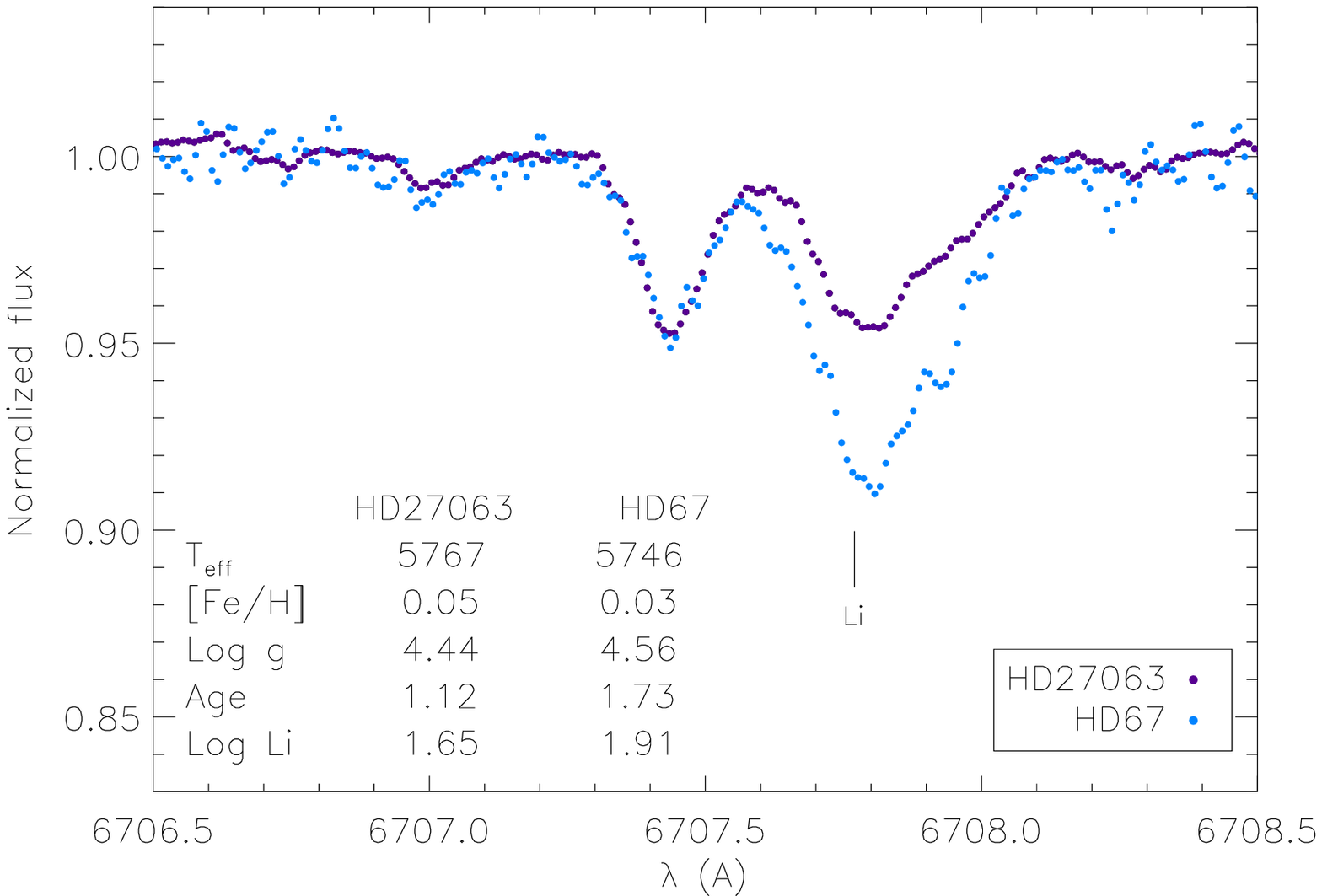}
\includegraphics[width=0.32\linewidth]{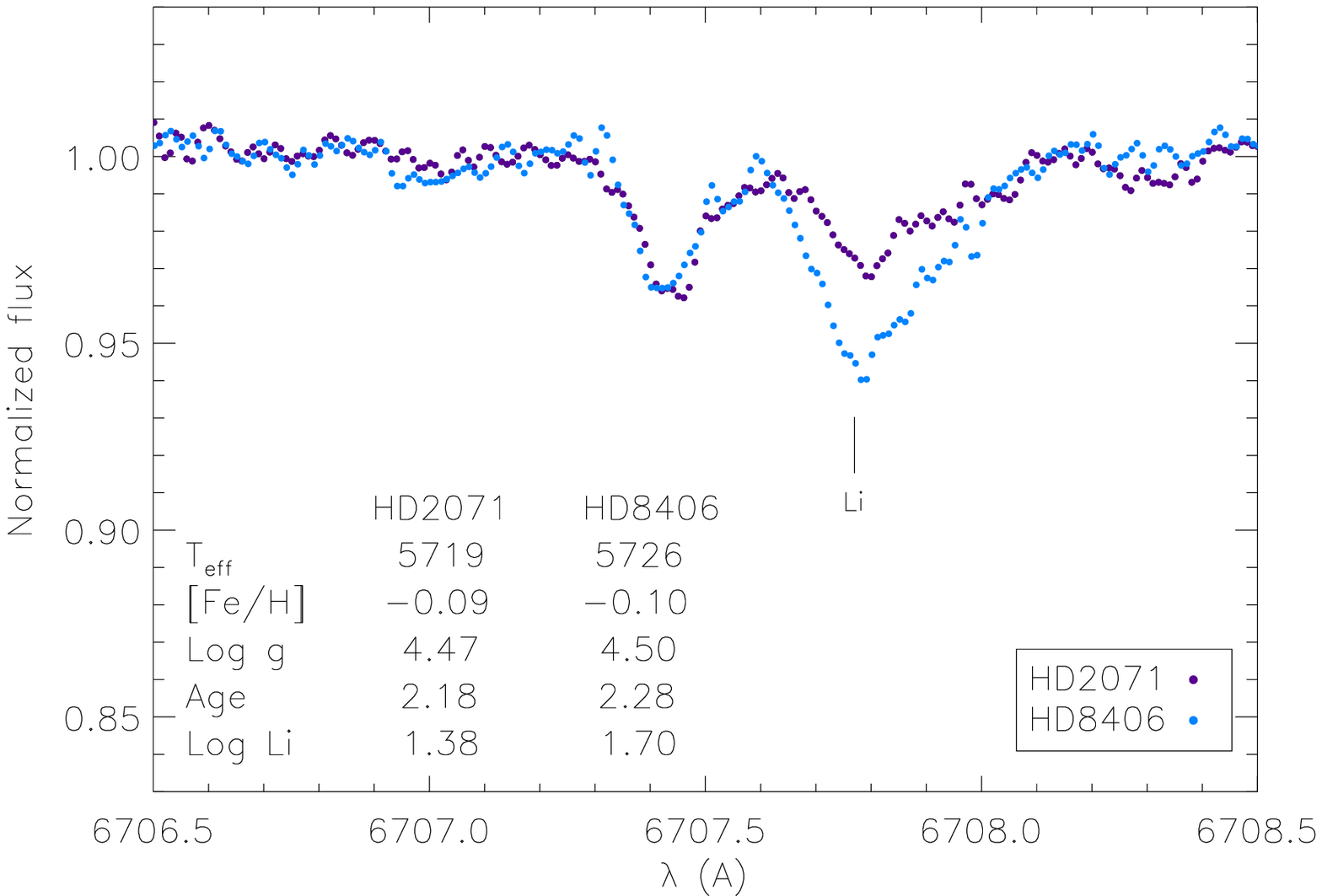}
\includegraphics[width=0.32\linewidth]{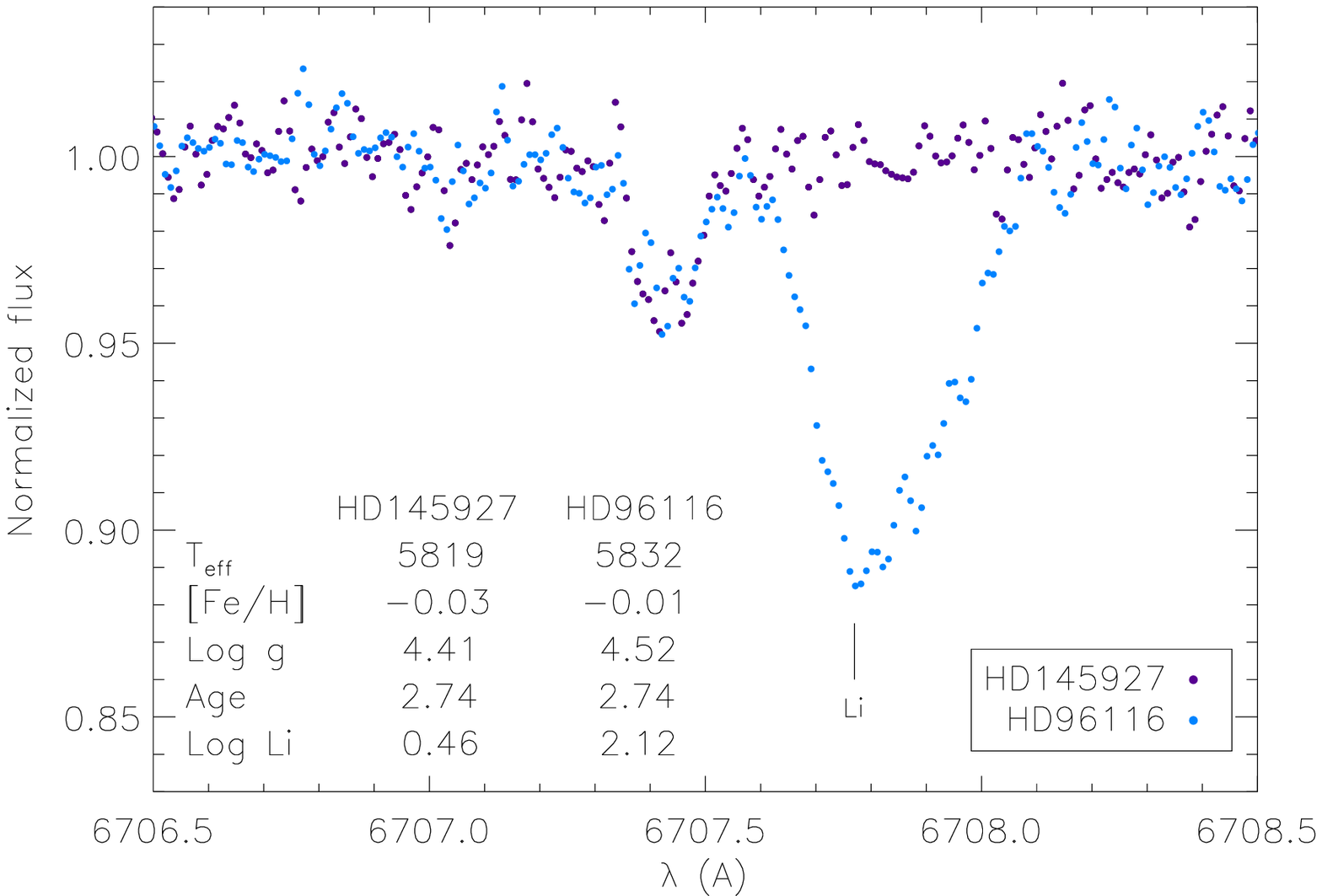}
\caption{Observed spectra for couples of stars with very similar parameters and different Li abundances. Planet hosts spectra are depicted with red dots and stars without known planets are depicted with blue or purple dots. For the couples with a planet host the minimum mass of the most massive planet in each system is indicated in the plot.}
\label{couples}

\end{figure*}

In Fig. \ref{couples} several couples of stars with very similar parameters are depicted. The stars in each pair have maximum differences of 40 K, 0.06 dex, 0.2 dex, 0.06 M$_{\odot}$ and 1 Gyr for \teff\ , [Fe/H], \logg\ , mass and age, respectively. These plots are a definitive demonstration that other parameter is affecting Li depletion, otherwise we would not be able to explain the difference in Li line for these almost twins stars (in each pair). The difference between parameters is about 1$\sigma$ while difference in Li abundances goes from 0.3 dex (more than 3$\sigma$) to more than 1.5 dex for the pair with HD 145927 and HD 96116 in which the two stars have the same age. We can find couples in which both stars do not have known planets (but present a quite distinct degree of Li depletion) and we can also find couples in which the planet host has a Li abundance higher than the single star. However, all those planet hosts (except HD 9446) have smaller planets (see next section). Indeed, it is easy to detect the Li absorption in some of those stars with smaller planets but when you move to planetary masses higher than that of Jupiter, the Li line almost dissapears. In any case, and regardless of the presence of planets, these huge differences in Li cannot be justified with the observed differences in stellar parameters.

Consequently, we can conclude that neither mass, nor age nor metallicity is the only cause of the extra Li depletion for our solar analogues. Other mechanisms such as rotationally induced mixing or overshooting mixing should be considered.


\subsection{Li and planetary parameters}
\begin{figure}
\centering
\includegraphics[width=9.0cm]{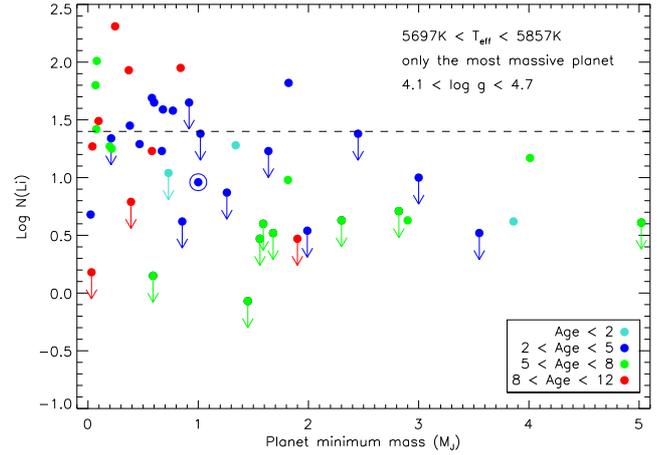}
\caption{Lithium abundances vs. minimum planetary masses for the most massive planet in each system. Colours denote different age ranges.} 
\label{massplan}
\end{figure}

We also searched for any possible relation of previously shown Li depletion with the physical and orbital parameters of the planets, that is, mass, period, eccentricity and semi-major axis. We did not find any dependence except possibly for the planetary mass. In Fig. \ref{massplan} we plot the Li abundances as a function of the more massive planet in each system (we note that using the sum of masses of all the planets in each system produces a similar plot). It seems that the destruction of Li is higher when the planet is more massive. This would make sense in a scenario where the disc is affecting the evolution of angular momentum and hence mixing mechanisms \citep{bouvier08}, since we could expect a higher effect if the disc is more massive and has a longer lifetime, conditions needed to form a giant planet. In fact, all the stars with planets more massive than Jupiter (except HD 9446) have depleted their Li significantly. Furthermore, the accretion processes are expected to be more frequent and violent when there is a giant planet in the disc and, as a consequence, produce Li destruction either by the increase of temperature in the base of the convective envelope \citep{baraffe10} or by extra-mixing triggered by thermohaline convection \citep{theado12}.\\

In Fig. \ref{massplan} we only constrain the \teff\ of the stars but not their \logg\ or [Fe/H] so some of the planet hosts with higher Li abundances have low \logg\ and might be slightly evolved. In fact, the planet hosts with higher Li detections are older than 5 Gyr, and half of them are older than 8 Gyr. As discussed in previous sections this might be the reason for the high Li abundance and not the mass of the planet, but what is sure is that all the stars with planets within solar \logg\ range and high Li abundances host less massive planets. Therefore, the mass of the disc in those systems would not be enough to affect the stellar angular momentum evolution or to produce intense accretion episodes and thus, those stars would present a different degree of Li depletion depending on other parameters. We could say that those planet hosts behave in a similar way as stars without planets, since some of them have destroyed their Li but others have not and the mechanism responsible to deplete it should be other than the presence of planets. We should also take into account that most of the Li abundances are upper limits so we can just conclude that Li is severely depleted at least when the planet is massive enough. In order to extract the information of upper limits, we computed the Kendall's tau correlation coefficient for censored data as done before. This test gives $\tau$ = 0.016 (with P=0.12), which means there is no correlation. If we consider only the detections $\tau$ decreases to -0.215 (with P=0.87), pointing to a possible correlation but the result is not significant enough. Nevertheless, we think that this plot shows an interesting relation and should be taken into account in further studies of Li abundances when the number of low mass planet hosts increases.

\section{Conclusions}

We present Li abundances in a large sample of planet hosts and "single" stars in the effective temperature range 5600\,K\,$<$\,\teff\,$<$\,5900\,K. 42 planet hosts and 235 "single" stars come from HARPS GTO sample while other 49 planet hosts have been observed with other instruments. We find that the average Li abundance for "single" stars is greater than for the planet hosts. This difference is more obvious when we look at solar twins with \teff\ = T$_{\odot} \pm$ 80 K, \logg\ = \logg$_{\odot}$ $\pm$ 0.2 and [Fe/H] = [Fe/H]$_{\odot}$ $\pm$ 0.2 where only 18\% of planet hosts show detections of A(Li) $>$ 1.4 whereas for "single" stars this value reaches to 48\% confirming previous studies \citep[e.g.][]{israelian09,gonzalez_li10,takeda10,sousa_li}.\\  

We check whether or not our abundances depend on other parameters like mass, age or metallicity. 
We find that most of our stars at very high metallicities present low Li content regardless of the presence of planets. On the other hand, if planet hosts were much older than the "single" stars we might expect that to be the reason of the extra depletion since Li shows a small decrease with age as observed in open clusters of several ages \citep{sestito05}. After removing the younger objects (age $<$ 1.5 Gyr) we show that planet hosts are around 1 Gyr older on average, but this small age difference cannot be responsible for the high difference in Li abundances (more than 3$\sigma$) found between stars with and without planets. However, some statistical tests including the Li upper limits, show that both populations might be not so different depending on the ranges in age and [Fe/H] used to define the subsamples.
On the other hand,
we show several examples of couples formed by twins 
with huge differences in abundance, up to 1.6 dex. Therefore, we suggest that the presence of giant planets (among other possible causes) might affect the Li content in solar twins.\\

The dispersion of Li abundances in stars of similar \teff\, mass, age and metallicity have also been reported in several open clusters like M67, which represents a good comparison for our observation since it has solar age and metallicity. Moreover, the standard model for stellar evolution, which considers only convection as mixing mechanism, cannot explain the extra Li depletion found in old stars (regardless of the presence of planets) during the MS. Therefore, there must exist any non-standard processes that produce this effect and we propose that the presence of planets is one of them. For instance, the stars with low Li in M67 might have planets which have not been discovered yet. Indeed, two of the planet candidates proposed in this cluster by \citet{pasquini12} have \teff\ in our solar range and present low Li abundances \citep{pasquini08}.
However we also find stars without detected planets which are heavily depleted in Li, hence, other mechanisms such as 
rotationally induced mixing \citep{pinsonneault90} must also be considered.
In fact, this kind of mixing is related to the initial rotation rates and angular momentum evolution of the star and thus has a strong connection with protoplanetary discs. Therefore, the presence of a planetary disc may be one of the causes to produce this kind of mixing. In addition, planets can also cause an extra Li depletion through other mechanisms like violent accretion-burst episodes of planetary material. 
Both mechanisms produce a higher effect when the protoplanetary discs are more massive, in agreement with our finding that Li is always destroyed if the star hosts a giant planet. 

\begin{acknowledgements}
E.D.M, S.G.S and V.Zh.A. acknowledge the support from the Funda\c{c}\~ao para a Ci\^encia e Tecnologia, FCT (Portugal) in the form of the fellowships SFRH/BPD/76606/2011, SFRH/BPD/47611/2008 and SFRH/BPD/70574/2010 from the FCT (Portugal), respectively. G.I. and J.I.G.H. acknowledge financial support from the Spanish Ministry project MINECO AYA2011-29060, and J.I.G.H. also from the Spanish Ministry of Economy and Competitiveness (MINECO) under the 2011 Severo Ochoa Program MINECO SEV-2011-0187. E.D.M, S.G.S., N.C.S. A.M. and V.Zh.A. thank for the support by the European Research Council/European Community under the FP7 through Starting Grant agreement number 239953, as well as the support through programme Ci\^encia\,2007 funded by FCT/MCTES (Portugal) and POPH/FSE (EC), and in the form of grant PTDC/CTE-AST/098528/2008.

We thank Pedro Figueira and the referee for useful comments and suggestions that helped improve the paper.

This research has made use of the SIMBAD database operated at CDS, Strasbourg (France) and the Encyclopaedia of Extrasolar Planets. This work has also made use of the IRAF facility.

\end{acknowledgements}

\bibliographystyle{aa}
\bibliography{edm_bibliography}

\begin{table*}
 \caption{Li abundances for stars with planets from HARPS GTO samples. Parameters from
\citet{sousa08,sousa_harps4,sousa_harps2}. The minimum mass of the most massive planet in the system is indicated in the last column (taken from The Extrasolar Planets Encyclopaedia, \citet{schneider}).}
\centering
\begin{tabular}{lcccrrrrcc}
\hline
\hline
\noalign{\smallskip}
Star & $T_\mathrm{eff}$ & $\log{g}$ & $\xi_t$ & [Fe/H]  & Age & Mass & ${\rm A(Li)}$ & Error & M \textit{sin} i\\  
     & (K)  &  (cm\,s$^{-2}$) & (km\,s$^{-1}$) &  & (Gyr)& (M$_{\odot}$) &  &  & (M$_{J}$)\\
\noalign{\smallskip}
\hline    
\noalign{\smallskip}
HARPS-1  & & & & & & & & &\\
\noalign{\smallskip}
\hline    
\noalign{\smallskip}
   \object{HD\, 1461\tablefootmark{a}} & 5765 &  4.38 &  0.97 &  0.19  &  3.73  &  1.05  &   0.68 &  0.10 &  0.024\\
   \object{HD\, 4308\tablefootmark{a}} & 5644 &  4.38 &  0.90 & -0.34 & 11.72  &  0.82  &  0.96 &  0.06 &  0.041\\
  \object{HD\, 16141} & 5806 &  4.19 &  1.11 &  0.16  &  6.55  &  1.07  &  1.25 &  0.10&  0.215\\
  \object{HD\, 16417\tablefootmark{a}} & 5841 &  4.16 &  1.18 &  0.13  &  5.99  &  1.12  & 1.80 &  0.04 &  0.069 \\
  \object{HD\, 20782} & 5774 &  4.37 &  1.00 & -0.06 &  9.21  &  0.94  &  $<$ 0.47 &   --  &  1.900 \\
  \object{HD\, 28185} & 5667 &  4.42 &  0.94 &  0.21 &  3.38  &  1.02  &  $<$ 0.26 &   --  &  5.700 \\
  \object{HD\, 31527\tablefootmark{a}} & 5898 &  4.45 &  1.09 & -0.17  &  7.57  &  0.95  &  1.91 &  0.03&  0.052 \\
  \object{HD\, 38858\tablefootmark{a}} & 5733 &  4.51 &  0.94 & -0.22  &  8.48  &  0.89  &  1.49 &  0.05&  0.096 \\
  \object{HD\, 45184\tablefootmark{a}} & 5869 &  4.47 &  1.03 &  0.04  &  2.28  &  1.05  &  2.06 &  0.04&  0.040 \\
  \object{HD\, 47186} & 5675 &  4.36 &  0.93 &  0.23 &  4.57  &  1.02  &      0.58 &  0.10 &  0.351  \\
  \object{HD\, 65216} & 5612 &  4.44 &  0.78 & -0.17 &  6.20  &  0.89  &      1.23 &  0.05 &  1.210 \\
  \object{HD\, 66428} & 5705 &  4.31 &  0.96 &  0.25 &  5.26  &  1.03  &  $<$ 0.71 &   --  &  2.820 \\
  \object{HD\, 70642} & 5668 &  4.40 &  0.82 &  0.18 &  2.69  &  1.02  &  $<$ 0.46 &   --  &  2.000 \\
  \object{HD\, 92788} & 5744 &  4.39 &  0.95 &  0.27 &  1.81  &  1.05  &      0.62 &  0.07 &  3.860 \\
  \object{HD\, 96700\tablefootmark{a}} & 5845 &  4.39 &  1.04 & -0.18  & 10.53  &  0.93  &   1.27 &  0.08 &  0.040\\
 \object{HD\, 102117} & 5657 &  4.31 &  0.99 &  0.28 &  8.40  &  1.01  &      0.52 &  0.10  &  0.172\\
 \object{HD\, 102365\tablefootmark{a}} & 5629 &  4.44 &  0.91 & -0.29  & 11.32  &  0.85  &  $<$ 0.30 &   --  &  0.050\\
 \object{HD\, 107148} & 5805 &  4.40 &  0.93 &  0.31 &  3.63  &  1.07  &  $<$ 1.34 &   --  &  0.210 \\
 \object{HD\, 114729} & 5844 &  4.19 &  1.23 & -0.28 & 11.00  &  0.95  &      1.95 &  0.04 &  0.840  \\
 \object{HD\, 117207} & 5667 &  4.32 &  1.01 &  0.22 &  6.31  &  1.01  &  $<$ 0.12 &   --  &  2.060 \\
 \object{HD\, 134987} & 5740 &  4.30 &  1.08 &  0.25 &  6.29  &  1.04  &  $<$ 0.60 &   --  &  1.590 \\
 \object{HD\, 141937} & 5893 &  4.45 &  1.00 &  0.13 &  1.08  &  1.09  &      2.35 &  0.04 &  9.700 \\
 \object{HD\, 147513} & 5858 &  4.50 &  1.03 &  0.03 &  0.64  &  1.05  &      2.05 &  0.05 &  1.210 \\
 \object{HD\, 160691} & 5780 &  4.27 &  1.09 &  0.30 &  6.36  &  1.07  &      0.98 &  0.10 &  1.814 \\
 \object{HD\, 190647} & 5639 &  4.18 &  0.99 &  0.23 &  8.47  &  1.04  &  $<$ 0.51 &   --  &  1.900 \\
 \object{HD\, 202206} & 5757 &  4.47 &  1.01 &  0.29 &  1.97  &  1.06  &      1.37 &  0.05 & 17.400  \\
 \object{HD\, 204313} & 5776 &  4.38 &  1.00 &  0.18 &  3.03  &  1.05  &  $<$ 0.52 &   --  &  3.550  \\
 \object{HD\, 222582} & 5779 &  4.37 &  1.00 & -0.01 &  7.48  &  0.97  &      0.85 &  0.15 &  7.750 \\
 \object{HD\, 126525} & 5638 &  4.37 &  0.90 & -0.10  &  9.60  &  0.89  &  $<$ -0.01 &   --  &  0.224\\
 \object{HD\, 134606} & 5633 &  4.38 &  1.00 &  0.27  &  7.09  &  0.99  &  $<$  0.39 &   --  &  0.121\\
 \object{HD\, 136352\tablefootmark{a}} & 5664 &  4.39 &  0.90 & -0.34  & 11.73  &  0.84  &  $<$ -0.10  & -- &  0.036\\
 \object{HD\, 150433} & 5665 &  4.43 &  0.88 & -0.36  & 11.43  &  0.82  &  $<$  0.27 &   --  &  0.168\\
 \object{HD\, 189567\tablefootmark{a}} & 5726 &  4.41 &  0.95 & -0.24  & 11.55  &  0.87  &  $<$  0.18 &   -- &  0.032 \\
 \object{HD\, 215456} & 5789 &  4.10 &  1.19 & -0.09  &  8.37  &  1.05  &       2.31 &  0.04&  0.246\\
\noalign{\smallskip}							      
\hline
\noalign{\smallskip}
HARPS-4  & & & & & & & & &\\
\noalign{\smallskip}
\hline    
\noalign{\smallskip}							      
 \object{HD\, 171028} & 5671 &  3.84 &  1.24 & -0.48  &  --   &  --   &  $<$  0.04  &   --  &  1.980\\
 \object{HD\, 181720} & 5792 &  4.25 &  1.16 & -0.53  & 11.22 &  0.92 &       1.93  &  0.02 &  0.370  \\
\noalign{\smallskip}							      
\hline
\noalign{\smallskip}
HARPS-2  & & & & & & & & &\\
\noalign{\smallskip}
\hline    
\noalign{\smallskip}							      
   \object{HD\, 6718} &  5723 &  4.44  &  0.84 &  -0.07   &  6.54 &  0.95  &  $<$  0.47&   --  &  1.560  \\
  \object{HD\, 28254} &  5653 &  4.15  &  1.08 &   0.36   &  7.75 &  1.06  &  $<$  0.59&   --  &  1.160  \\
  \object{HD\, 30177} &  5601 &  4.34  &  0.89 &   0.37   &  5.66 &  0.99  &  $<$  0.51&   --  &  7.700 \\
  \object{HD\, 44219} &  5766 &  4.20  &  1.06 &   0.04   &  8.43 &  1.01  &       1.23&  0.10 &  0.580\\
 \object{HD\, 109271\tablefootmark{a}} &  5783 &  4.28  &  0.97 &   0.10   &  7.30 &  1.05  &  1.42&  0.08 &  0.076 \\
 \object{HD\, 207832} &  5718 &  4.45  &  0.86 &   0.15   &  1.90 &  1.01  &  $<$  1.04&   --  &  0.730\\
\noalign{\smallskip}
\hline
\noalign{\smallskip}
\end{tabular}
\begin{flushleft}
\tablefoottext{a}{Stars which only host Neptune and Earth-like planets.}\\
\end{flushleft}
\label{tabla_harps_plan}
\end{table*}   

\clearpage

\begin{table*}
\label{tabla_harps_comp}
\caption{Li abundances for stars without detected planets from HARPS GTO samples. Parameters from
\citet{sousa08,sousa_harps4,sousa_harps2}. Complete table in online version.}
\centering
\begin{tabular}{lcccrrrrc}
\hline
\hline
\noalign{\smallskip} 
Star & $T_\mathrm{eff}$ & $\log{g}$ & $\xi_t$ & [Fe/H]  & Age  & Mass & ${\rm A(Li)}$ & Error \\  
     & (K)  &  (cm\,s$^{-2}$) & (km\,s$^{-1}$) &  & (Gyr) & (M$_{\odot}$) &  &\\
\noalign{\smallskip} 
\hline
\noalign{\smallskip} 
HARPS-1  & & & & & & & & \\
\noalign{\smallskip}
\hline    
\noalign{\smallskip}
    \object{HD\, 1320}  & 5679 &  4.49  &  0.85  & -0.27  &    7.76  &   0.86  &       1.29  &  0.06  \\
    \object{HD\, 2071}  & 5719 &  4.47  &  0.95  & -0.09  &    2.18  &   0.97  &       1.38  &  0.07  \\
    \object{HD\, 4307}  & 5812 &  4.10  &  1.22  & -0.23  &    8.68  &   1.03  &       2.39  &  0.04  \\
    \object{HD\, 4915}  & 5658 &  4.52  &  0.90  & -0.21  &    3.12  &   0.91  &       1.38  &  0.05  \\
    \object{HD\, 8406}  & 5726 &  4.50  &  0.87  & -0.10  &    2.28  &   0.97  &       1.70  &  0.05  \\
   \object{HD\, 11505}  & 5752 &  4.38  &  0.99  & -0.22  &   11.43  &   0.88  &  $<$  0.35  &   --   \\
   \object{HD\, 12387}  & 5700 &  4.39  &  0.93  & -0.24  &   11.56  &   0.86  &  $<$  0.15  &   --   \\
\noalign{\smallskip}
\hline    
\end{tabular}
\end{table*}

\clearpage

\begin{table*}
\caption{Planet hosts stars not belonging to the HARPS-GTO sample. The minimum mass of the most massive planet in each system is taken from The Extrasolar Planets Encyclopaedia, \citet{schneider}).}
\centering
\begin{tabular}{lcccrrrrcccc}
\hline
\hline
\noalign{\smallskip}
Star & $T_\mathrm{eff}$ & $\log{g}$ & $\xi_t$ & [Fe/H] & Age & Mass  & ${\rm A(Li)}$ & Error& M \textit{sin} i & Flag & Reference\\  
     & (K)  &  (cm\,s$^{-2}$) & (km\,s$^{-1}$) &  & (Gyr) & (M$_{\odot}$) &  & & (M$_{J}$) & &\\
\noalign{\smallskip}
\hline    
\noalign{\smallskip}
\object{HD \,45350} &  5650 &  4.29 &  1.08 &  0.28 &    8.22 &    1.01 &  $<$ 0.52 &   --  &  1.790&  [6] &  this work \\
\object{HD \,79498} &  5814 &  4.40 &  1.09 &  0.24 &    1.10 &    1.08 &      1.28 &  0.10 &  1.340&  [6] &  this work \\
\object{HD \,179079\tablefootmark{a}} & 5742 & 4.11 & 1.22 & 0.26 & 6.86 & 1.09 & 2.01 & 0.05 &  0.080&  [6] &  this work  \\
\object{HIP \,14810} &  5601 &  4.43 &  0.96 &  0.26 &    4.35 &    0.99 &  $<$ 0.84 &   --  & 3.880 &   [1] &  this work  \\  
\noalign{\smallskip}
\hline 
\noalign{\smallskip}
\object{HD \,4113} &  5688 & 4.40 &  1.08 &  0.20 &  5.39 & 1.01 &  $<$ 0.86 &   --  &  1.560 & [5] & \citet{tamuz08} \\
\object{HD \,4203} &  5636 & 4.23 &  1.12 &  0.40 &  8.23 & 1.01 &  $<$ 0.65 &   --  &  2.070 & [3] & \citet{santos04}\\
\object{HD \,6434} &  5835 & 4.60 &  1.53 & -0.52 & 10.21 & 0.86 &  $<$ 0.79 &   --  &  0.390 & [3]  & \citet{santos04}  \\
\object{HD \,12661} &  5702 & 4.33 &  1.05 &  0.36 &  5.84 & 1.02 &  $<$ 0.63 &   --  &  2.300 & [8]  & \citet{santos04}  \\
\object{HD \,49674} &  5644 & 4.37 &  0.89 &  0.33 &  2.13 & 1.01 &  $<$ 1.00 &   --  &  0.115 & [4]  & \citet{santos04}  \\
\object{HD \,73526} &  5699 & 4.27 &  1.26 &  0.27 &  7.28 & 1.05 &      0.63 &  0.10 &  2.900 & [1]  & \citet{santos04}  \\
\object{HD \,76700} &  5737 & 4.25 &  1.18 &  0.41 &  6.80 & 1.05 &      1.27 &  0.10 &  0.190 & [1]  & \citet{santos04}  \\
\object{HD \,106252} &  5899 & 4.34 &  1.08 & -0.01 &  3.52 & 1.03 &      1.71 &  0.05 &  7.560 & [1]  & \citet{santos04} \\ 
\object{HD \,109749} &  5899 & 4.31 &  1.13 &  0.32 &  2.98 & 1.12 &      2.24 &  0.05 &  0.280 & [5]  & \citet{sousa06} \\
\object{HD \,114762} &  5884 & 4.22 &  1.31 & -0.70 & 11.61 & 0.87 &      2.07 &  0.03 & 10.980 & [1]  & \citet{santos04} \\ 
\object{HD \,154857} &  5610 & 4.02 &  1.30 & -0.23 &  4.52 & 1.22 &      1.68 &  0.03 &  1.800 & [3]  & \citet{santos05} \\ 
\object{HD \,168443} &  5617 & 4.22 &  1.21 &  0.06 &  9.68 & 1.01 &  $<$ 0.45 &   --  & 17.193 & [4]  & \citet{santos04} \\ 
\object{HD \,178911B} &  5600 & 4.44 &  0.95 &  0.27 &  5.38 & 0.99 &  $<$ 0.94 &   --  &  6.292 & [4]  & \citet{santos04} \\ 
\object{HD \,186427} &  5772 & 4.40 &  1.07 &  0.08 &  6.23 & 1.00 &  $<$ 0.52 &   --  &  1.680 & [4]  & \citet{santos04} \\ 
\object{HD \,187123} &  5845 & 4.42 &  1.10 &  0.13 &  4.53 & 1.05 &  $<$ 0.54 &   --  &  1.990 & [4]  & \citet{santos04} \\ 
\object{HD \,188015} &  5793 & 4.49 &  1.14 &  0.30 &  2.22 & 1.07 &  $<$ 0.87 &   --  &  1.260 & [3]  & \citet{santos05} \\ 
\object{HD \,195019} &  5859 & 4.32 &  1.27 &  0.09 &  6.53 & 1.07 &      1.53 &  0.06 &  3.700 & [8]  & \citet{santos04} \\ 
\object{HD \,216437} &  5887 & 4.30 &  1.31 &  0.25 &  4.65 & 1.15 &      1.99 &  0.06 &  1.820 & [1]  & \citet{santos04} \\ 
\object{HD \,217014} &  5804 & 4.42 &  1.20 &  0.20 &  3.94 & 1.06 &      1.29 &  0.10 &  0.468 & [3]  & \citet{santos04} \\ 
\object{HD \,217107} &  5645 & 4.31 &  1.06 &  0.37 &  7.30 & 0.99 &  $<$ 0.61 &   --  &  2.490 & [3]  & \citet{santos04} \\ 
\noalign{\smallskip}
\hline 
\noalign{\smallskip}                    
   \object{HD \,9446} &  5793 & 4.53 &  1.01 &  0.09 &  2.49 & 1.03 &      1.82 &  0.05 &  1.820 & [7] & \citet{hebrard10} \\
 \object{HD \,109246} &  5844 & 4.46 &  1.01 &  0.10 &  2.08 & 1.04 &      1.58 &  0.10 &  0.770 & [7] & \citet{boisse10}  \\ 
   \object{Kepler-17} &  5781 & 4.53 &  1.73 &  0.26 &  3.68 & 1.07 &  $<$ 1.38 &   --  &  2.450 & [7] & \citet{bonomo12}  \\
   \object{  KOI-204} &  5757 & 4.15 &  1.75 &  0.26 &  4.03 & 1.06 &  $<$ 1.38 &   --  &  1.020 & [7] & \citet{bonomo12}  \\
   \object{     XO-1} &  5754 & 4.61 &  1.07 & -0.01 &  --   &  --  &      1.33 &  0.12 &  0.900 & [4] & \citet{ammler09} \\
\noalign{\smallskip}
\hline                
\noalign{\smallskip}
  \object{HD \,23127} &  5891 & 4.23 & 1.26 &  0.41  & 4.32  & 1.20  &     2.67 &  0.02 &  1.500 & [1] & \citet{santos13}  \\
  \object{HD \,24040} &  5840 & 4.30 & 1.14 &  0.20  & 5.36  & 1.09  &     1.17 &  0.10 &  4.010 & [1] & \citet{santos13}  \\
  \object{HD \,27631} &  5700 & 4.37 & 1.00 & -0.11  & 5.99  & 0.93  & $<$-0.07 &   --  &  1.450 & [3] & \citet{santos13}  \\
  \object{HD \,96167} &  5823 & 4.16 & 1.28 &  0.38  & 4.47  & 1.22  &     1.59 &  0.08 &  0.680 & [3] & \citet{santos13}  \\
  \object{HD \,98649} &  5714 & 4.37 & 1.01 & -0.03  & 4.24  & 0.96  & $<$-0.54 &   --  &  6.800 & [3] & \citet{santos13}  \\
 \object{HD \,126614} &  5601 & 4.25 & 1.17 &  0.50  & 8.57  & 0.99  & $<$ 0.57 &   --  &  0.380 & [1] & \citet{santos13}  \\
 \object{HD \,129445} &  5646 & 4.28 & 1.14 &  0.37  & 4.95  & 1.00  & $<$ 0.77 &   --  &  1.600 & [3] & \citet{santos13}  \\
 \object{HD \,152079} &  5785 & 4.38 & 1.09 &  0.29  & 2.98  & 1.06  & $<$ 1.00 &   --  &  3.000 & [2] & \citet{santos13}  \\
 \object{HD \,154672} &  5743 & 4.27 & 1.08 &  0.25  & 6.94  & 1.06  & $<$ 0.61 &   --  &  5.020 & [1] & \citet{santos13}  \\
 \object{HD \,170469} &  5845 & 4.28 & 1.17 &  0.30  & 4.54  & 1.09  &     1.23 &  0.10 &  0.670 & [1] & \citet{santos13}  \\
\noalign{\smallskip}
\hline                
\noalign{\smallskip}
\object{  CoRoT-9} & 5613 & 4.35 &  0.90 & -0.02 & 5.15 & 0.94 & $<$ 0.88 &  --  & 0.840 & [2] & \citet{mortier13_transits} \\
\object{ CoRoT-12} & 5715 & 4.66 &  1.07 &  0.17 & 4.31 & 1.01 & $<$ 1.65 &  --  &  0.917& [2] & \citet{mortier13_transits} \\
\object{   WASP-5} & 5785 & 4.54 &  0.96 &  0.17 & 3.34 & 1.04 & $<$ 1.23 &  --  &  1.637& [1] & \citet{mortier13_transits} \\
\object{   WASP-8} & 5690 & 4.42 &  1.25 &  0.29 & 3.22 & 1.03 &     1.54 & 0.15 &  2.244& [3] & \citet{mortier13_transits} \\
\object{  WASP-16} & 5726 & 4.34 &  0.97 &  0.13 & 4.43 & 1.02 & $<$ 0.62 &  --  &  0.855& [2] & \citet{mortier13_transits} \\
\object{  WASP-25} & 5736 & 4.52 &  1.11 &  0.06 & 3.94 & 1.00 &     1.69 & 0.05 &  0.580& [2] & \citet{mortier13_transits} \\
\object{  WASP-34} & 5704 & 4.35 &  0.97 &  0.08 & 5.15 & 0.99 & $<$ 0.15 &  --  &  0.590& [3] & \citet{mortier13_transits} \\
\object{  WASP-56} & 5797 & 4.44 &  1.19 &  0.43 & 3.01 & 1.07 &     1.65 & 0.10 &  0.600& [3] & \citet{mortier13_transits} \\
\object{  WASP-63} & 5715 & 4.29 &  1.28 &  0.28 & 4.36 & 1.04 &     1.45 & 0.07 &  0.380& [3] & \citet{mortier13_transits} \\
\object{ WASP-77A} & 5605 & 4.37 &  1.09 &  0.07 & 5.55 & 0.95 & $<$ 0.59 &  --  &  1.760& [3] & \citet{mortier13_transits} \\
\noalign{\smallskip}                                                                                                 
\hline                                                                                                                

\end{tabular}
\\
\begin{flushleft}
\tablefoot{ Flag: [1] UVES; [2] HARPS; [3] FEROS; [4] SARG; 
[5] CORALIE; [6] NOT;  [7] SOPHIE; [8] UES}\\
\tablefoottext{a}{Stars which only host Neptune and Earth-like planets.}\\
\end{flushleft}
\label{tab_otros}
\end{table*}      

\clearpage




\end{document}